\documentclass{aastex631}

\usepackage{graphicx}
\usepackage[utf8]{inputenc}

\usepackage{amsmath}
\usepackage{color,soul}

\usepackage{gensymb}
\usepackage{longtable}
\usepackage{array}
\usepackage{amssymb}

\begin{document}


\graphicspath{{./}{figures/}}


\title{Correcting Exoplanet Transmission Spectra for Stellar Activity with an Optimised Retrieval Framework}
\author{Alexandra Thompson}
\affiliation{Department of Physics and Astronomy, University College London, Gower Street, WC1E 6BT London, United Kingdom}
\author{Alfredo Biagini}
\affiliation{University of Palermo, Department of Physics and Chemistry `Emilio Segrè' Via Archirafi, 36, 90123 Palermo, Italy)}
\affiliation{INAF-Osservatorio Astronomico di Palermo, Piazza del Parlamento, 1, 90134 Palermo, Italy}
\author{Gianluca Cracchiolo}
\affiliation{University of Palermo, Department of Physics and Chemistry `Emilio Segrè' Via Archirafi, 36, 90123 Palermo, Italy)}
\affiliation{INAF-Osservatorio Astronomico di Palermo, Piazza del Parlamento, 1, 90134 Palermo, Italy}
\author{Antonino Petralia}
\affiliation{INAF-Osservatorio Astronomico di Palermo, Piazza del Parlamento, 1, 90134 Palermo, Italy}
\author{Quentin Changeat}
\affiliation{European Space Agency (ESA)
ESA Office, Space Telescope Science Institute (STScI), 3700 San Martin Drive, Baltimore MD 21218, United States of America}
\affiliation{Department of Physics and Astronomy, University College London, Gower Street, WC1E 6BT London, United Kingdom}
\author{Arianna Saba}
\affiliation{Department of Physics and Astronomy, University College London, Gower Street, WC1E 6BT London, United Kingdom}
\author{Giuseppe Morello}
\affiliation{Instituto de Astrof\'isica de Canarias (IAC), 38205 La Laguna, Tenerife, Spain}
\affiliation{Departamento de Astrof\'isica, Universidad de La Laguna (ULL), 38206, La Laguna, Tenerife, Spain}
\affiliation{Department of Space, Earth and Environment, Chalmers University of Technology, SE-412 96 Gothenburg, Sweden}
\author{Mario Morvan}
\affiliation{Department of Physics and Astronomy, University College London, Gower Street, WC1E 6BT London, United Kingdom}
\author{Giuseppina Micela}
\affiliation{INAF-Osservatorio Astronomico di Palermo, Piazza del Parlamento, 1, 90134 Palermo, Italy}
\author{Giovanna Tinetti}
\affiliation{Department of Physics and Astronomy, University College London, Gower Street, WC1E 6BT London, United Kingdom}


\begin{abstract}
The chromatic contamination that arises from photospheric heterogeneities e.g. spots and faculae on the host star presents a significant noise source for exoplanet transmission spectra. If this contamination is not corrected for, it can introduce substantial bias in our analysis of the planetary atmosphere. We utilise two stellar models of differing complexity, \texttt{StARPA} and \texttt{ASteRA}, to explore the biases introduced by stellar contamination in retrieval under differing degrees of stellar activity. We use the retrieval framework TauREx3 and a grid of 27 synthetic, spot-contaminated transmission spectra to investigate potential biases and to determine how complex our stellar models must be in order to accurately extract the planetary parameters from transmission spectra. The input observation is generated using the more complex model (\texttt{StARPA}), in which the spot latitude is an additional, fixable parameter. This observation is then fed into a combined stellar-planetary retrieval which contains a simplified stellar model (\texttt{ASteRA}). Our results confirm that the inclusion of stellar activity parameters in retrieval minimises bias under all activity regimes considered. \texttt{ASteRA} performs very well under low to moderate activity conditions, retrieving the planetary parameters with a high degree of accuracy. For the most active cases, characterised by larger, higher temperature contrast spots, some minor residual bias remains due to \texttt{ASteRA} neglecting the interplay between the spot and the limb darkening effect. As a result of this, we find larger errors in retrieved planetary parameters for central spots (0\degree) and those found close to the limb (60\degree) than those at intermediate latitudes (30\degree). 

\end{abstract}

\section{Introduction}
\label{introduction}

With over 5000 confirmed exoplanet detections, and this number rapidly increasing with the contribution of ground-based surveys e.g. HARPS \citep{mayor2003} and space-based missions such as TESS \citep{ricker2014}, we are entering a period of unprecedented potential for exoplanet characterisation. Present exoplanet observations and analyses are laying the foundations for the large-scale population studies that will be conducted with next generation space observatories such as JWST \citep{bean2018} and, in less than a decade, Ariel \citep{tinetti2021}. Planets from multiple, distinct regions of the known parameter space have already been analysed in detail with transmission \citep[e.g.][]{charbonneau2002, tinetti2007, sing2016, tsiaras2018, hoeijmakers2019, pinhas2019, tsiaras2019, anisman2020, pluriel2020, skaf2020, edwards2022, gressier2022, rustamkulov2022, saba2022}, emission \citep[e.g.][]{swain2008, crouzet2014, evans2017, mikalevans2020, changeat2022b} and phase curve spectroscopy \citep[e.g.][]{knutson2012, stevenson2014, arcangeli2019, feng2020, irwin2020, vonessen2020, changeat2022, dang2022}. All of these techniques come with the caveat that the planet and star are observed as an unresolved source, with the host star providing the light source that makes these methods possible. As such, disentangling the stellar signals from the planetary ones is a challenging, but essential part of any exoplanet characterisation pipeline. The objective of this paper is to outline a simple, scalable method of disentangling these signals that can be implemented seamlessly in retrievals. The primary aim is to accurately retrieve the planetary parameters in the presence of stellar activity. We investigate what biases the simplified model assumptions could produce as a result of missing physics and how limiting these biases could be. As a secondary aim, we also explore what useful information about the host star can be extracted from a combined stellar-planetary retrieval.

Our current understanding of exoplanet host stars indicates that a substantial fraction of them will display moderate to high levels of activity. As such, stellar contamination will likely be one of the most dominant noise sources in exoplanetary observations. Evidence for this is shown in the form of activity indicators e.g. Ca H-K lines, S-index etc. \citep{gomesdasilva2011, cauley2018, klein2022}, variability amplitudes from long-term photometric monitoring e.g. with Kepler \citep{ciardi2011, mcquillan2014} and the many spot and plage-crossing events that have been detected in light-curves of transiting exoplanets \citep[e.g.][]{pont2013, oshagh2014, morris2017, espinoza2019}. With higher resolution observations rapidly becoming available to us it will be essential to account for this activity, ideally in a homogeneous way, in order to characterise exoplanet atmospheres with the high precision we are aiming for and subsequently to conduct larger, comparative population studies. Observations covering larger wavelength ranges will also be highly beneficial as they allow for an easier identification and subsequent correction for stellar contamination.

Some trends in activity have already been well documented in the literature, with higher levels of activity typically observed for later type K-M dwarfs \citep{goulding2012, jackson2013, mcquillan2014} as stars transition towards becoming fully convective \citep{reiners2010} and also in the case of fast rotating, young stars or young stellar objects (YSOs) \citep[e.g.][]{gullysantiago2017, jarvinen2018, morris2020, klein2022}. These scenarios are particularly important as much of the pioneering exoplanetary science is now focusing on planets orbiting these stars. These smaller, later-type stars are frequently observed to host small planets \citep{dressing2015}, which are crucial for pushing the limits of our current characterisation techniques and, ultimately for answering questions surrounding potential habitability. Contamination due to stellar activity has the potential to negatively affect observations of all types of planets, albeit in different ways. For larger planets with H-dominated, primary atmospheres, the bias introduced by not accounting for contamination could potentially result in inaccurate retrieved atmospheric parameters such as chemical abundances and temperature-pressure profiles e.g. \citep{saba2022}. These inaccuracies may subsequently be propagated into discussions surrounding other key research areas e.g. elemental ratios \citep{oberg2011, pacetti2022} and disequilibrium chemistry \citep{venot2020} leading to an incorrect interpretation of the planet as a whole. For small planets, particularly those possessing secondary atmospheres, atmospheric absorption features will generally tend to be weaker, making them much more susceptible to being obscured or altered by stellar contamination \citep[e.g.][]{ballerini2012, rackham2018, zhang2018}. In recent years the number of observations of exoplanets orbiting YSOs / young main sequence stars e.g. AU Mic \citep{szabo2021, klein2022}, V1298 Tau \citep{feinstein2022, maggio2022} and K2-33b \citep{thao2023}, has also increased as, despite being strongly affected by stellar activity, studying these systems provides us with a unique opportunity to begin to fill in gaps in our understanding of the stages of planetary formation and evolution \citep{raymond2022}.

In transmission spectroscopy an active star is capable of contaminating the observed spectrum in multiple ways by causing the transit chord to differ from the disk-integrated stellar spectrum. The most efficient methods of modelling and correcting for this contamination have been and are still being explored extensively \citep[e.g.][]{czesla2009, garcia2010, silvavalio2010, sing2011, ballerini2012, oshagh2014, micela2015, herrero2016, newton2016, rackham2018, bixel2019, cracchiolo2021a, cracchiolo2021b}. At the spectral intervals observed for exoplanet characterisation, typically the optical and near infrared (NIR) regimes, we are observing the stellar photosphere. On the photosphere, magnetic activity manifests in the form of heterogeneities such as cooler spots and hotter faculae \citep{solanki2003, berdyugina2005}. If present, these features are the dominant sources of stellar contamination in this wavelength range. The strength of this contamination and its overall effect on the observed spectrum differs depending on what type of active regions are present and whether or not they are occulted by the transiting planet. Occulted active regions are arguably easier to identify due to the characteristic bumps they introduce in the light-curves. These bumps, whilst not always easily corrected for, can often be masked/removed from the light-curve. Unocculted features are more insidious as they affect the transit depth of the entire light-curve, without imparting any obvious identifying feature on it. In addition to this, unocculted features could potentially be far more common, as the planet only occults a small fraction of the stellar disk as it transits.

Unocculted spots, the focus of this study, reduce the average flux that originates from the regions of the star not crossed by the planet. Their presence introduces a wavelength dependent signal that deepens the transit light-curve, resulting in overestimates of the planetary radius, particularly in the optical regime. Stellar contamination is highly chromatic, with its effects appearing substantially stronger and more evident at shorter wavelengths \citep[e.g.][]{ballerini2012, rackham2018}. The main concern when contamination is present but not corrected for is that it may introduce biases in the retrieved planetary parameters. It is capable of both obscuring absorption features in the case of unocculted faculae, or mimicking/strengthening them in the case of unocculted spots. An additional complication arises in that stellar contamination is temporally variable. Modulation occurs predominantly on the timescale of stellar rotation but also through spot evolution and longer timescale magnetic cycles \citep{ciardi2011, bradshaw2014, mcquillan2014, zhang2018}. This means that each observation of each exoplanetary system should ideally be corrected individually before they can accurately be combined or analysed simultaneously. This limitation poses a problem for small planets in particular, as we will need to stack observations from multiple visits to obtain a high enough SNR for a successful retrieval analysis.

For the above reasons, stellar activity is one of the most pressing challenges to the accurate characterisation of exoplanetary atmospheres at present. Modelling the star as a more complex astrophysical body, rather than as a homogeneous light source, is essential in tackling this issue. Despite this, increases in model complexity are not always beneficial and care needs to be taken when introducing additional parameters in retrievals as this can have detrimental effects. Using models with a higher dimensionality than is necessary may increase the risk of overfitting the data or injecting a bias through the model choice. Alternatively, if many of the parameters are degenerate, the retrieval may not be able to converge on a solution at all. More complex models will also intrinsically come with an associated computational cost. The aim of this paper is to find the middle ground by determining a stellar model that encompasses all of the essential physics of stellar activity required to remove any potential biases, but without overcomplicating the model to the detriment of retrieval reliability or computing time. To achieve this we utilise two approaches to modelling stellar activity in the form of a single, unocculted spot. The predominant difference between them is the consideration of the limb darkening effect, or the lack thereof. \texttt{StARPA}, the more complex of the two models, encompasses the interplay of the presence of spots and the limb darkening effect by fixing the spot position on the stellar disk. The \texttt{StARPA} model is then used as the forward model in a benchmarking retrieval exercise which uses \texttt{ASteRA}, the simpler of the two models, in which this interplay between spot contamination and limb darkening is neglected and all parameters describing the spot and the planetary atmosphere are fitted simultaneously.Stellar activity has been considered and fit for in several previous retrieval studies \citep{pinhas2018, bixel2019, espinoza2019, edwards2021}. In order to obtain accurate planetary parameters and maximise the information content from our retrievals we need to have a good understanding of what the potential biases from stellar activity are. These biases can originate both from neglecting stellar contamination or could be introduced or incompletely removed by our chosen correction process. Retrieval biases due to not correcting for stellar activity have previously been systematically explored in \citet{iyer2020}. This work presents the first investigation into the potential residual biases that are left behind as a result of neglecting the limb darkening effect in the correction process. A similar investigation was conducted in \citet{cracchiolo2021b}, albeit on a smaller scale.
In Section \ref{experiment} we introduce our grid of 27 spot-contaminated stellar disks from which we generate the contaminated transmission spectra. This grid is produced using the \texttt{StARPA }model where the activity is characterised by a single unocculted starspot. This spot is parameterised by its temperature contrast with respect to the quiet photosphere, its radius and the latitude of the spot centre. The \texttt{StARPA} model is described in detail in Section \ref{StARPA} followed by a description of the simplified \texttt{ASteRA} model that is used in retrievals in Section \ref{ASteRA}. We conduct retrievals on the grid of contaminated spectra to investigate under which conditions \texttt{ASteRA} is capable of accurately retrieving the planetary, and to a lesser extent the stellar parameters, in the presence of varying degrees of stellar contamination. The results of these retrievals are given in Section \ref{results}. Detailed analysis of these results alongside initial investigations into more complex, realistic cases involving multiple spots and spots that are separated into an umbra and penumbra are given in Section \ref{discussion}. In this section we also discuss when using a simplified stellar model, such as \texttt{ASteRA}, is valid at first order and under what activity conditions the assumptions within it begin to break down. Our final, concluding remarks are given in Section \ref{conclusion}.

\section{Methodology}

This section introduces the overall experiment framework (Section \ref{experiment}) and the two modelling approaches utilised throughout the work (Sections \ref{StARPA} and \ref{ASteRA}) to explore the effect of model complexity on the accuracy of the retrieved planetary and, to a lesser extent, stellar parameters. The main objective of this work is to develop a stellar activity correction method that is both computationally fast and capable of retrieving planetary parameters accurately for host stars displaying different levels of activity. It is of course desirable to also be able to accurately retrieve the stellar parameters, however, we prioritise the planetary parameters as these are fundamental for the large-scale characterisation of exoplanet populations intended to be carried out with ongoing and future missions e.g. JWST and Ariel. We aim to determine under which activity conditions using a simplified model is sufficient, and, if/where it is insufficient, investigate any potential biases it may introduce. 

These questions surrounding model complexity are essential questions to answer as, as our underlying physical models become more realistic, their dimensionality increases drastically, which will likely require an increase in computation time and, in the worst scenario can be detrimental to retrieval accuracy. On the contrary, by using an oversimplified model we risk neglecting underlying physical processes that are necessary in order to fully and accurately interpret our observations. As such, we are increasingly facing a compromise between complexity and the computing cost such models require, although state of the art machine learning methods could potentially mitigate this \citep[e.g.][]{yip2022}.

To explore the ability of the \texttt{ASteRA} model to accurately retrieve the planetary parameters, we generate a grid of 27 spot-contaminated transmission spectra as forward models. These are produced using the \texttt{StARPA} model (Section \ref{StARPA}) and are then used as input observations in retrievals where the stellar contamination is accounted for in a combined stellar-planetary retrieval, using \texttt{TauREx3} and the \texttt{ASteRA} plugin described in Section \ref{ASteRA}. We also introduce a case with an uncontaminated, quiet star, termed ‘Case 0’, which acts as a baseline for the subsequent retrieval analysis. Each contaminated stellar disk in our grid is characterised by a single, unocculted spot which is itself parameterised by its temperature contrast ($\Delta T_{Spot}$) i.e. how much cooler than the quiescent photosphere the spot is, the spot radius ($R_{Spot}$) normalised to the stellar radius and the latitude of the spot centre ($\phi_{Spot}$) from which we can probe the effect of limb darkening. These 27 cases are discussed in greater detail in Section \ref{experiment}.

\subsection{Simulating the Uncontaminated Planetary Transmission Spectrum for a Synthetic Star-Planet System}
\label{uncontaminated}

To determine how extreme the contamination effects are for each spot case considered and explore how well these effects can be mitigated we must first produce a synthetic star-planet system, for which the stellar and planetary parameters are known a priori. We consider a synthetic K-dwarf host star characterised by the following parameters ($T_{eff}$ = 4750 K, $R_{\star}$ = 0.8 $R_{\odot}$, $M_{\star}$ = 0.8 $M_{\odot}$) and displaying activity in the form of a single unocculted spot that we model using the methodology described in Section \ref{StARPA}. This is the same methodology outlined in \citet{cracchiolo2021b} with some subsequent, minor improvements to its efficiency. A K-dwarf was chosen as stars of later spectral types are typically more likely to be active \citep{berdyugina2005, ciardi2011, hartman2011, mcquillan2014, rackham2018, rackham2019}. We decided against using an M-dwarf host for this preliminary study as this is the region of the main sequence in which stars transition to a fully convective regime \citep{reiners2010}, as such, stellar activity may manifest differently on these stars. We choose to focus on single spot cases for this initial benchmarking study for several reasons. Firstly, this represents the simplest spot-contaminated disk model that can be considered which makes it invaluable as a baseline for future investigations. Encompassed within this is that it reduces the dimensionality required for the input model. Having multiple spots present requires the location of each of them on the stellar disk to be defined in order to accurately compute the interplay between their properties and the limb darkening effect. Secondly, a single, larger spot also allows us to probe the extremes of this interplay in a way that multiple spots will not as, for a given filling factor, the active region is concentrated at a single latitude rather than dispersed over multiple locations on the stellar disk. Despite this, the extension to multiple spot cases is comparatively straightforward as described in \citet{cracchiolo2021b}. The results of several preliminary multiple spot cases are presented in Section \ref{multispot} for completeness.

The transiting planet is a temperate sub-Neptune ($R_{P}$ = 3 $R_{\oplus}$ (0.273 $R_{Jup}$), $M_{P}$= 5 $M_{\oplus}$ (0.0157 $M_{Jup}$) and $T_{P}$ = 400 K) with a primary atmosphere containing water ($log (H_{2}O)$ = -3) and H and He present as fill gases with a ratio of 0.172. Rayleigh scattering and collision induced absorption (CIA) are also included and introduce wavelength dependent contributions to the opacity. A smaller planet with a primary atmosphere is used as its scale height should result in detectable absorption features but the smaller signal to noise ratio (SNR) of such features means that they are more susceptible to being masked by stellar contamination. As such, the accuracy of our correction method, and any biases that may be inadvertently introduced, are comparatively far more important here than when considering an atmosphere with a much higher SNR, for example that of a hot Jupiter. The orbital inclination is set to 88$\degree$ so that the effects of a central unocculted spot ($\phi_{spot}$ = 0$\degree$) can also be explored.

Using the stellar and planetary parameters described above and the retrieval code \texttt{TauREx3} \citep{alrefaie2021, alrefaie2022}, we produce a forward model that is equivalent to the idealised, uncontaminated transmission spectrum that would be observed in the presence of a completely homogeneous host star. This synthetic spectrum has a wavelength coverage of 0.5 – 9.5 $\mu m$ and a resolution of 200. This wavelength range has been chosen as it is similar to the regions that are/will be covered by the JWST and Ariel instruments but with a greater coverage and resolution in the optical regime where the effects of stellar contamination are strongest. Error bars of 10 ppm are introduced for all data points regardless of wavelength. This noise level was chosen as 10 ppm is broadly consistent with the most precise observations obtained with the Hubble Space Telescope (HST) instruments\citep[e.g.][]{edwards2022}, albeit for larger planets further into the IR. We acknowledge that obtaining this level of precision in reality would be extremely challenging and heavily reliant on an accurate characterisation of the host star to mitigate the astrophysical noise. We rationalise our choice of using small error bars as our goal at this stage is to be limited by the model assumptions/limitations and not the instrument performance, which will be a focus of future work. Testing our correction methods on an idealised case first allows us to quantify how effective they are and ensure that they do not introduce any unknown, intrinsic bias before using them with real observations. 

\subsection{The Experiment Framework: 27 Spot-Contaminated Cases}
\label{experiment}

\begin{figure}[htp]
    \centering
    \includegraphics[width=0.55\textwidth, height=0.92\textheight]{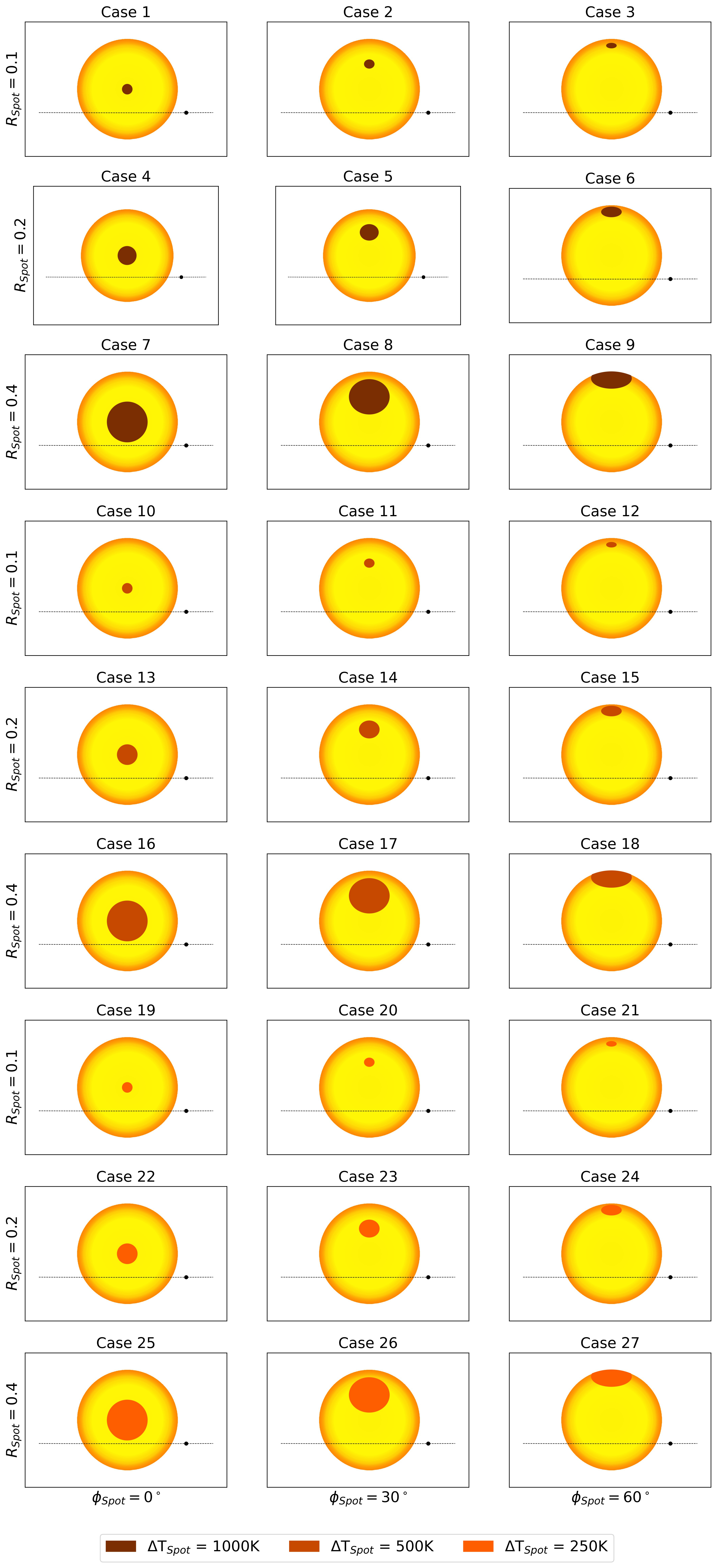}
    \caption{Visual representations of the 27 spotted star cases investigated in this study. The spot colour corresponds to its temperature contrast with respect to the quiescent photosphere which has a temperature $T_{Phot}$ = 4750 K.}
    \label{fig:27cases}
\end{figure}

Throughout this study, we consider 27 spot-contaminated cases (and an uncontaminated case as a reference frame) for the star-planet system outlined in the previous section. From this grid we investigate the potential biases that an unocculted spot may introduce and how these may vary as a function of the three spot parameters considered ($\Delta T_{Spot}$, $R_{Spot}$ and $\phi_{Spot}$). The contaminated stellar disks are produced with \texttt{StARPA} (Section \ref{StARPA}) and the following values are considered for each spot parameter: $\Delta T_{Spot}$ = [250, 500, 1000] K, $R_{Spot}$ = [0.1, 0.2, 0.4] $R_{\star}$ and $\phi_{Spot}$ = [0, 30, 60]$\degree$. This results in 27 unique parameter combinations in which only one spot parameter is varied at a time. The full set of spot parameters used in each case are given in Table~\ref{tab:27cases}. Visual representations of each case are shown in Fig~\ref{fig:27cases}.

\begin{table}[htp]
\centering
\caption{An outline of the 27 single spot scenarios considered in this study. Each case is characterised by a unique combination of three spot parameters: the temperature contrast with respect to the photosphere ($\Delta T_{Spot}$), the spot radius normalised to the stellar radius ($R_{Spot}$) and the latitude of the spot centre ($\phi_{Spot}$). The spot filling factor ($F_{Spot}$) is calculated assuming an elliptical projection onto the surface when the spot centre is at non-zero latitudes. An uncontaminated case is also considered to act as a control case (Case 0).} 
\label{tab:27cases} 
\begin{tabular}{ccccc}
\hline

Case No. & $\Delta T_{Spot}$ \ (K) & $R_{Spot}$ ($R_{*}$) &  $\phi_{Spot}$ \ (\degree) & $F_{Spot}$ ($\%$) \\ \hline \hline
0 & N/A & N/A & N/A & N/A\\
1 & 1000 & 0.1 & 0 & 1\\
2 & 1000 & 0.1 & 30 & 0.867\\
3 & 1000 & 0.1 & 60 & 0.498\\
4 & 1000 & 0.2 & 0 & 4\\
5 & 1000 & 0.2 & 30 & 3.462\\
6 & 1000 & 0.2 & 60 & 2.001\\
7 & 1000 & 0.4 & 0 & 16\\
8 & 1000 & 0.4 & 30 & 13.858\\
9 & 1000 & 0.4 & 60 & 7.997\\
10 & 500 & 0.1 & 0 & 1\\
11 & 500 & 0.1 & 30 & 0.867\\
12 & 500 & 0.1 & 60 & 0.498\\
13 & 500 & 0.2 & 0 & 4\\
14 & 500 & 0.2 & 30 & 3.462\\
15 & 500 & 0.2 & 60 & 2.001\\
16 & 500 & 0.4 & 0 & 16\\
17 & 500 & 0.4 & 30 & 13.858\\
18 & 500 & 0.4 & 60 & 7.997\\
19 & 250 & 0.1 & 0 & 1\\
20 & 250 & 0.1 & 30 & 0.867\\
21 & 250 & 0.1 & 60 & 0.498\\
22 & 250 & 0.2 & 0 & 4\\
23 & 250 & 0.2 & 30 & 3.462\\
24 & 250 & 0.2 & 60 & 2.001\\
25 & 250 & 0.4 & 0 & 16\\
26 & 250 & 0.4 & 30 & 13.858\\
27 & 250 & 0.4 & 60 & 7.997\\ \hline

\end{tabular}
\end{table}

\clearpage

\subsection{The Input Stellar Model: \texttt{StARPA} (Stellar Activity Removal for Planetary Atmospheres)}
\label{StARPA}

\begin{figure}[h!]
    \centering
    \includegraphics[width=0.95\textwidth, height=0.25\textheight]{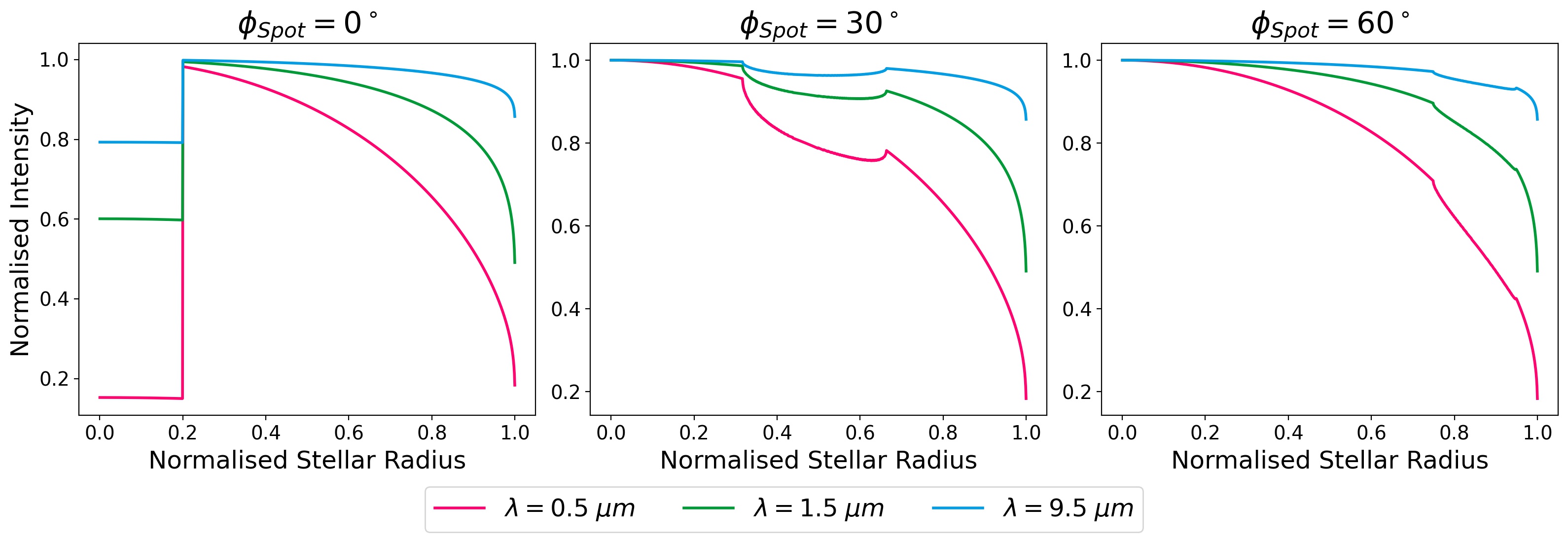}
    \caption{Intensity profiles, normalised to the flux emitted from the disk centre of a homogeneous star with $T_{eff}$=4750 K, for a spot-contaminated star of the same temperature possessing a 0.2$R_{*}$, $\Delta T_{Spot}$=1000 K spot located at latitudes of 0\degree (left), 30\degree (centre) and 60\degree (right) and viewed at wavelengths of 0.5 $\mu m$ (pink), 1.5 $\mu m$ (green) and 9.5 $\mu m$ (blue) respectively.}
    \label{fig:spotty_norm_int}
\end{figure}

From the uncontaminated planetary transit spectrum produced with \texttt{TauREx}, the grid of 27 spot-contaminated spectra described above is constructed using the \citet{cracchiolo2021b} model with some subsequent improvements made to the implementation. The key point of this methodology with respect to this study is that, through accounting for the limb darkening effect and defining the exact position of the spot on the stellar disk, any interactions between contamination and limb darkening can be explored. As in \citet{cracchiolo2021b}, the spots are modelled as being circular but with an elliptical 2D projection on the visible stellar disk at non zero latitudes. The out-of-transit stellar flux from the spot-contaminated star is computed using quadrature integration; the star is divided into 1000 equally spaced annuli and the fractions of each of these annuli covered by the spot are calculated. This enables us to compute both the absolute and normalised intensity profiles (Fig.~\ref{fig:spotty_norm_int}).

To create the contaminated stellar disks we model the contribution of the quiet photosphere and the spot using the BT-Settl \citep{allard2012, baraffe2015} stellar spectral model grids from the PHOENIX library \citep{husser2013}. The spectral emission densities (SEDs) corresponding to the photosphere and the spot are governed by three fundamental stellar parameters: the stellar effective temperature ($T_{eff}$), the stellar metallicity $[M/H]$ and the stellar surface gravity ($\log g$). For the purposes of this study the metallicity and gravity are fixed at $[M/H]$=0 (solar metallicity) and $\log g$=4.5 respectively, in order to isolate the effects of active regions with contrasting temperatures. These are reasonable assumptions for low resolution spectroscopy but may need to be reconsidered at higher resolution. In particular, the treatment of $\log g$ may need improvement, as it has been suggested that spots may be characterised by a lower $\log g$ than that of the photosphere due to the localised increase in magnetic pressure \citep[e.g.][]{solanki2003}. In total we require four SEDs, one corresponding to the photospheric temperature ($T_{Phot}$ = 4750 K) and three corresponding to the spot temperatures considered in this work: 3750 K, 4250 K and 4500 K, equivalent to $\Delta T_{spot}$ of 1000 K, 500 K and 250 K respectively. 

Limb darkening is a well-known phenomenon acting to reduce the flux originating from the limbs of the stellar disk with respect to its centre \citep{claret2000, howarth2011}. It also varies as a function of wavelength and as such, different bands are characterised by different intensity profiles with the strongest effects seen in the optical. To account for the limb darkening effect within our forward model we use the \texttt{ExoTETHyS} package, specifically the \texttt{SAIL} and \texttt{BOATS} subpackages \citep{morello2020a, morello2020b, morello2021}, to calculate the limb darkening coefficients (LDCs) for the star using the PHOENIX$\_$2012$\_$13 database \citep{claret2012, claret2013} of BT-Settl models and the Claret four-coefficient law \citep{claret2000}:

\begin{equation}
\label{limb_darkening}
  \frac{I_{\lambda}(\mu)}{I_{\lambda}(1)} = 1 - \sum_{n=1}^{4}a{_n,_\lambda} (1-\mu^{n/2}),  
\end{equation}

where $\lambda$ is the wavelength/bandpass being considered, $\mu$=cos$\theta$ (where $\theta$ is the angle between the line of sight and the normal at the stellar surface), $I_\lambda$($\mu$) is the stellar intensity profile, $I_\lambda$(1) is the intensity at the disk centre (i.e. where $\mu$=1) and $\alpha_{n,\lambda}$ are the LDCs.

\begin{figure}[h!]
    \centering
    \includegraphics[width=0.75\textwidth, height=0.4\textheight]{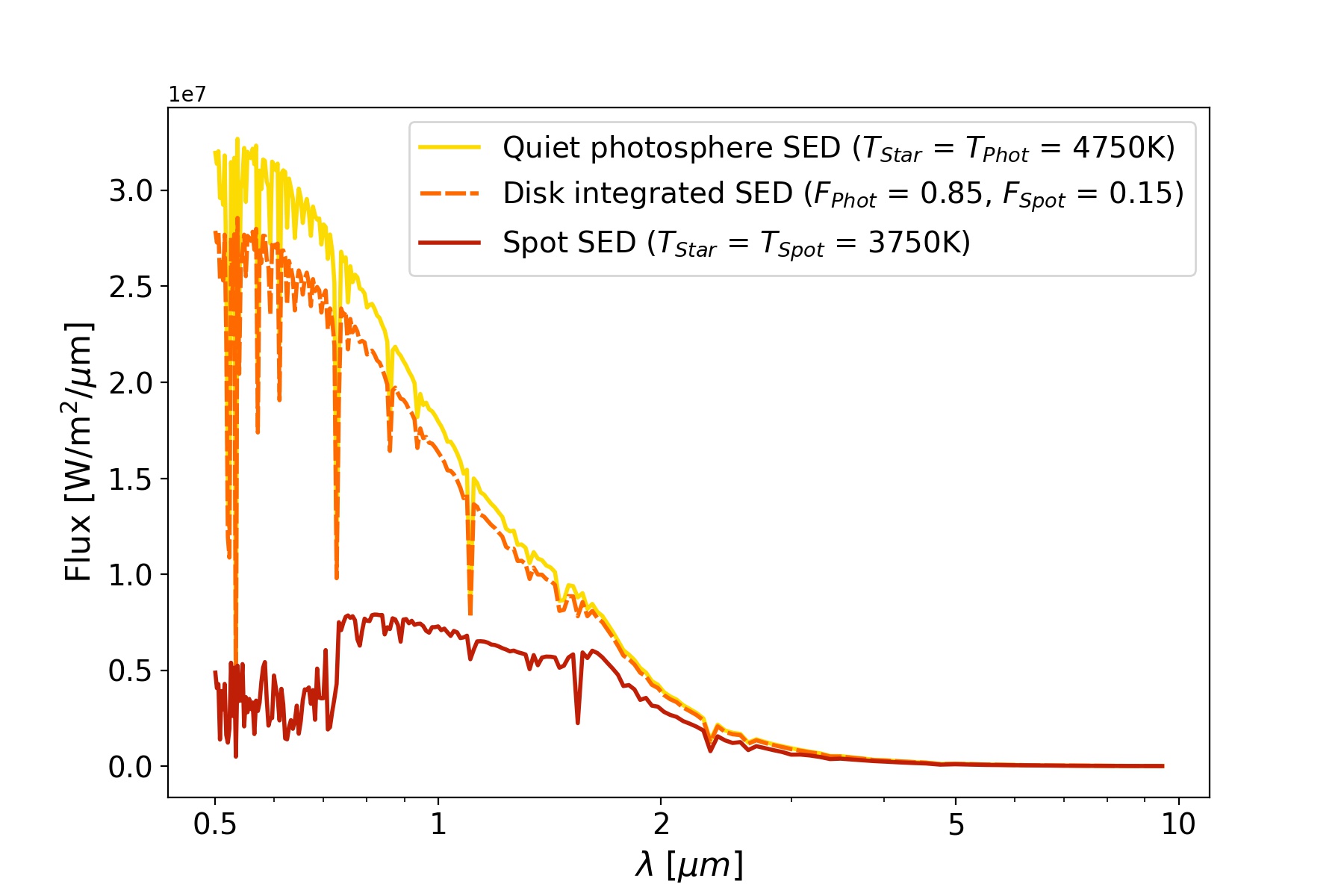}
    \caption{Three SEDs that are relevant to the construction of the forward stellar model. The two solid line SEDs correspond to the two temperature components that make up the surface of the active, heterogeneous stars considered in this work, the quiet photosphere, $T_{Phot}$=4750 K (yellow) and the cooler starspot, $T_{Spot}$=3750 K (red). The orange, dashed line SED represents the average, disk integrated SED that would be observed in accordance with Eq. \ref{SED_spoteq} due to the weighted contributions of the quiet photosphere and the spot, assuming a spot filling factor of $F_{Spot}$ = 0.15 (and therefore a $F_{Phot}$ = 0.85). We highlight that the spot filling factor is varied throughout the spot-contaminated, cases considered in this work (Table \ref{tab:27cases}), as such the disk averaged SEDs will vary between cases. 
}
    \label{fig:SEDs}
\end{figure}

We model the spot using the same limb darkening coefficients that have been calculated for the star which is a reasonable approximation at first order, but again, may be revised for higher resolution observations. The absolute fluxes of the star and the spot are also calculated using \texttt{ExoTETHyS} from which the normalised intensity profiles of the spotted star can be calculated as a function of wavelength (Fig.~\ref{fig:spotty_norm_int}). The emission from the spot-contaminated stellar disk is calculated as in Eq. \ref{SED_spoteq}, where $S_{Star, \lambda}$, $S_{Phot, \lambda}$ and $S_{Spot, \lambda}$ are the spectra of the average star, the quiescent photosphere and the spot respectively for a given wavelength ($\lambda$) and $F_{Spot}$ is the spot filling factor. As such the resulting spectrum for the active star is essentially a combination of the photosphere and spot SEDs weighted by their relative covering fractions (Fig.~\ref{fig:SEDs}). The limb darkening effect, which has already been defined for the 1000 annuli considered using Eq. \ref{limb_darkening}, is also accounted for in this stage.

\begin{equation}
\label{SED_spoteq}
    S_{Star, \lambda} = ((1-F_{Spot}) \times S_{Phot, \lambda}) + (F_{Spot} \times S_{Spot,\lambda})  ,
\end{equation}

The resulting chromatic contamination can be described as acting as a contamination factor ($\varepsilon$) relative to the nominal transit depth i.e. the uncontaminated spectrum that would be observed in the case of an inactive star (Eq. \ref{eq:chromatic}). Where contamination is present only in the form of lower temperature spots, as in this study, $\varepsilon_{\lambda}$ $>$ 1 resulting in increased transit depths at all wavelengths. 

\begin{equation}
\label{eq:chromatic}
    \varepsilon_{\lambda} = \frac{1}{1-F_{Spot}\left(1-\frac{S_{\lambda,Spot}}{S_{\lambda, Phot}}\right)} ,
\end{equation}

\begin{figure}[htp]
    \centering
    \includegraphics[width=0.95\textwidth, height=0.22\textheight]{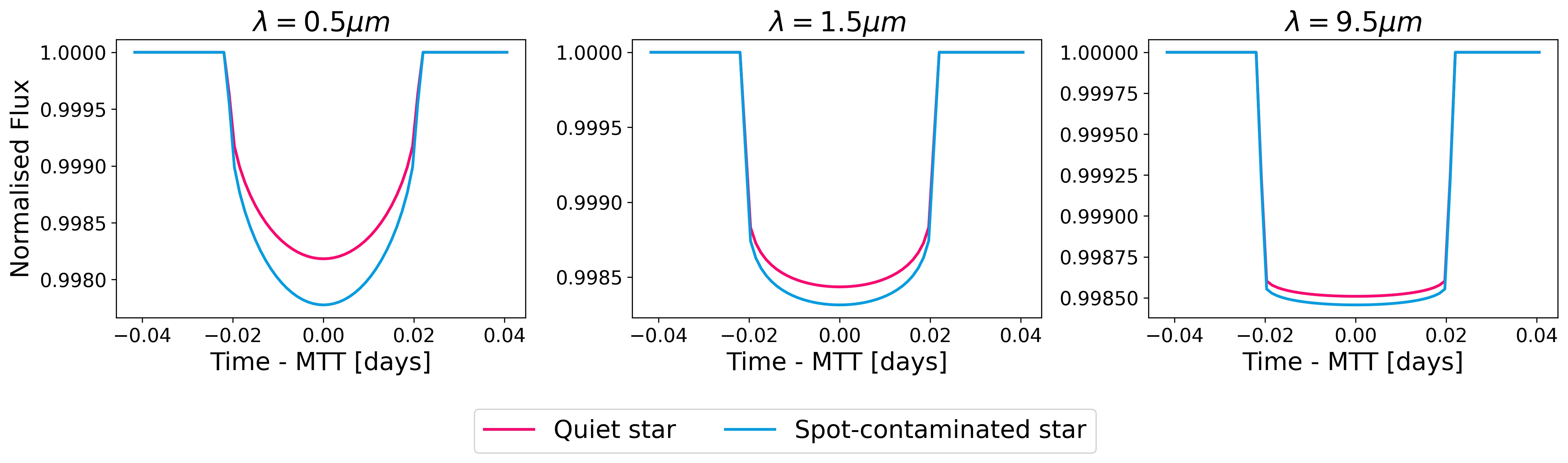}
    \caption{Uncontaminated (pink) and spot-contaminated (blue) light-curves computed at 0.5$\mu m$ (left), 1.5$\mu m$ (centre) and 9.5$\mu m$ (right) in the case of a large, central, high-contrast spot (Case 7 in Table~\ref{tab:27cases})}
    \label{fig:lightcurves}
\end{figure}

\begin{figure}[htp]
    \centering
    \includegraphics[width=0.8\textwidth, height=0.43\textheight]{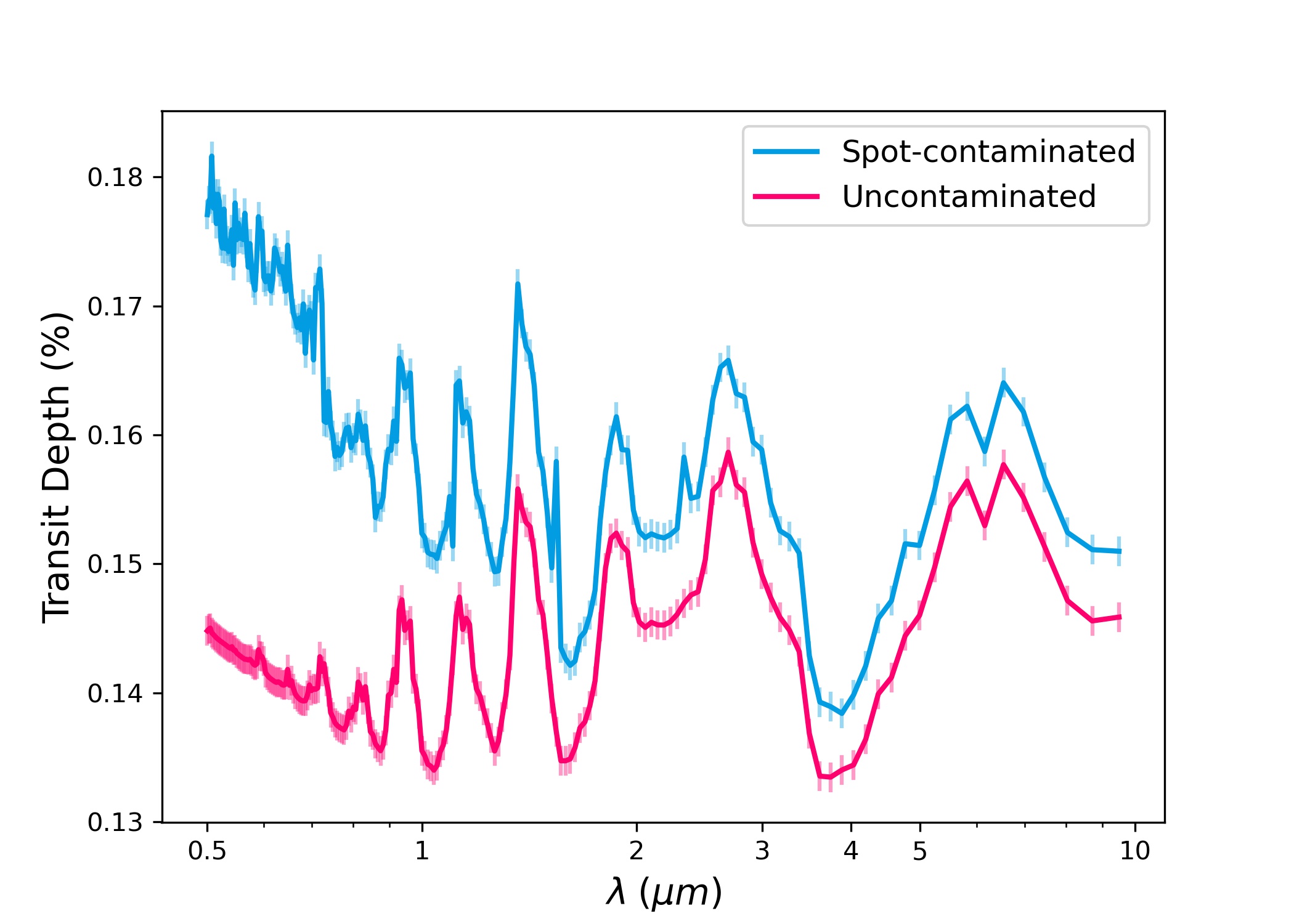}
    \caption{The effect of an extreme case of stellar contamination introduced by a large, high-contrast, unocculted, central spot (Case 7) on the observed transmission spectrum. The transit depths are overestimated at all wavelengths for the contaminated spectrum (blue) in comparison to the uncontaminated spectrum (pink). Error bars for both spectra are equivalent to 10 ppm. The magnitude of this overestimation increases exponentially as a function of decreasing wavelength with the strongest contamination seen in the optical regime.}
    \label{fig:contam_spec}
\end{figure}

From the constructed, spot-contaminated stellar disks, we then use the open source, light-curve analysis package \texttt{pylightcurve} \citep{tsiaras2016} to produce the spot-contaminated light-curves (Fig.~\ref{fig:lightcurves}) for all 200 wavelength intervals. These light-curves are then used to construct the contaminated transmission spectrum for each spot case. We highlight that for this preliminary investigation, as we are only aiming to construct the contaminated transmission spectra for use as inputs in the retrievals, we make the assumption that the orbital parameters used in \texttt{pylightcurve} are well constrained/known a priori and we do not introduce any Gaussian noise at the light-curve fitting stage to avoid introducing any additional bias that may arise arbitrarily. As in Section \ref{uncontaminated}, error bars of 10 ppm are assumed for the input spectra. A comparison of the uncontaminated transmission spectrum and the spectrum with the highest degree of stellar contamination considered in this study is shown in Fig.~\ref{fig:contam_spec}. The strongly wavelength-dependent nature of the spot contamination is evident, with it imparting a steep, positive bluewards slope on the spectrum at wavelengths shortwards of $\sim$ 2 $\mu m$. Longwards of 2 $\mu m$, the contamination acts to introduce an almost constant positive offset equivalent to an overestimate in the transit depth on the order of 5$\%$ ($>$ 50 ppm) for the worst case scenario.

\subsection{The Retrieval Stellar Model: \texttt{ASteRA} (Active Stellar Retrieval Algorithm)}
\label{ASteRA}

Retrievals on the grid of spot-contaminated spectra produced with the \texttt{StARPA} model are conducted using the fully Bayesian atmospheric retrieval code \texttt{TauREx3} \citep{alrefaie2021}. The main benefit of using \texttt{TauREx} for this study is that its modular design allows for a large amount of freedom and flexibility in testing and introducing new models. Mitigating for potential contamination due to stellar activity has been conducted in previous studies in the form of combined stellar-planetary retrievals \citep[e.g.][]{pinhas2018, bixel2019, espinoza2019} and hopefully this consideration of the host star will progressively become the new norm within the exoplanetary retrieval community. Whilst this is not the first study focused on retrieval biases
as a function of the degree of stellar heterogeneity, parameterised by temperature contrast and spatial extent \citep{iyer2020}, this is the first time that that bias has been explored as a function of three spot parameters, with the inclusion of their position on the stellar disk which governs how they interact with the limb darkening effect.

The \texttt{ASteRA} plugin introduces a new, heterogeneous star class to \texttt{TauREx} which follows a formalism similar to that used in previous stellar activity studies \citep[e.g.][]{rackham2018, rackham2019}. Instead of modelling the host star as being homogeneous and characterised by a single temperature/SED, the active star is modelled as a combination of multiple distinct temperature components, in this case two: the quiet photosphere and the spot. The spot and photosphere SEDs are homogeneous disk-integrated models as they are not spatially resolved. The observed disk-integrated stellar spectrum is then produced by combining the two SEDs, weighted by the filling factor of each component, as shown in Fig. \ref{fig:SEDs} and described in Eq. \ref{SED_spoteq}  

Using the heterogeneous star model requires the addition of two extra fitting parameters to a ‘normal’ retrieval; the spot temperature $T_{Spot}$ and its filling factor $F_{Spot}$. For the purposes of this study we restrict the active regions considered to spots only, however, \texttt{ASteRA} is easily extended to incorporate faculae, as has been done in the analysis of HST STIS observations of the hot Jupiter WASP-17b \citep{saba2022}.

The core difference between the stellar models of \texttt{StARPA} and \texttt{ASteRA} is the treatment of limb darkening, or lack thereof. In contrast to the forward model produced with \texttt{StARPA}, the \texttt{ASteRA} stellar model is simpler in that the interplay between the spot and the limb darkening effect is neglected, making it similar to the initial model used in \citet{cracchiolo2021a}. Conducting retrievals without the inclusion of limb darkening helps us to gauge how important its inclusion is for low resolution spectroscopy, particularly with respect to active host stars. With limb darkening neglected, the relative position of the spot on the stellar disk becomes unimportant provided that the entire spot is unocculted, reducing the dimensionality of the model by one. \texttt{ASteRA} requires the fundamental stellar parameters, i.e. effective temperature, metallicity and $\log g$ as inputs in order to select the correct, corresponding PHOENIX spectra and these are fixed within the retrieval, as are the orbital parameters. This is done with the assumption that for real observations, these parameters will be reasonably accurately and homogeneously constrained a priori. Indeed, this is a high priority and ongoing effort within the Ariel consortium regarding the stellar parameters \citep{danielski2022, magrini2022} and the ExoClock project for the orbital parameters \citep{kokori2022}.

For each spot case, two separate retrievals were conducted; one accounting for activity by fitting for $T_{Spot}$ and $F_{Spot}$ and one where the activity is not accounted for, despite being present. The planetary parameters $R_{P}$, $T_{P}$ and $log(H_{2}O)$ are fit for in all retrievals and the prior bounds set for each parameter are given in Table~\ref{tab:fitparams}. All retrievals conducted with \texttt{TauREx} use \texttt{MultiNest} \citep{feroz2009, buchner2014} as the sampler with 500 live points and an evidence tolerance of 0.5. The retrievals with and without activity have a dimensionality of five and three respectively. Conducting two retrievals allows for a comparison of the Bayesian evidence to obtain the Bayes factor which quantifies how strongly the model with activity is favoured over the one without. This provides an additional, statistically motivated way of quantifying how strong the spot contamination effects are. The retrieved values are then compared to the input/ground truth values for each parameter. From this we determine how accurately the stellar and planetary parameters can be retrieved using \texttt{ASteRA} and if any bias is introduced as a result. The results of these retrievals are given in Section \ref{results}.
\clearpage

\begin{table}[h]
\centering
\caption{Fitting parameters used within the retrievals with \texttt{ASteRA} and \texttt{TauREx} and their prior bounds. \smallskip} \label{tab:fitparams} 
\begin{tabular}{lll}
\hline
Fitted Parameter & Prior Bounds   & Scale  \\ \hline \hline
$R_{P}$ ($R_{Jup}$)        & 0.75$R_{P}$ ; 1.5$R_{P}$ & linear \\
$T_{P}$ (K)            & 100 ; 1000     & linear \\
$T_{Spot}$ (K)        & 3000 ; 4700    & linear \\
$F_{Spot}$            & 0 ; 0.99       & linear \\
$log (H_{2}O)$              & -12 ; -1       & log$_{10}$ \\ \hline
\end{tabular}
\end{table}

\section{Results}
\label{results}

\subsection{Retrievals Accounting for the Spot Contamination}
\label{results_spot}

In this section we present the results of the retrievals for the 27 spot cases outlined in Section \ref{experiment}. To account for the spot-contamination in retrieval, the spot parameters $T_{Spot}$ and $F_{Spot}$ are fitted for using the \texttt{ASteRA} plugin. In contrast, in Section \ref{results_nospot}, retrievals are conducted on the same contaminated spectra but with the incorrect assumption that the star is homogeneous.The same retrievals are also conducted on the uncontaminated transmission spectrum which we term `Case 0'. Case 0 acts as both a frame of reference for the highest precision and accuracy obtainable with \texttt{TauREx} for the simulated spectra used in this study, and as a verification that the use of \texttt{ASteRA} does not introduce any intrinsic bias. The results of these retrievals are given in Table~\ref{tab:retr_activity} and a visual representation of the retrieval accuracy with respect to the planetary parameters is given in Fig.~\ref{fig:retr_planet}. On inspection of the retrieved planetary parameters alone, the outlook is overall very positive, with the retrieved values falling very close to the ground truth in almost all of the cases considered. Intuitively, it becomes apparent that the largest errors in the planetary parameters are obtained for the cases where the spectra are most strongly contaminated. These cases are characterised by the largest, highest contrast spots considered in this study e.g. Cases 7, 8  and 9. Despite showing the largest errors in this study, we emphasise that these errors are not substantial enough to result in a large misinterpretation of the planet. The retrieved parameters are also substantially more accurate than those obtained when the stellar contamination is neglected in the retrieval, the results of which are presented in the following section (Section \ref{results_nospot}). One other thing that becomes particularly evident here is that \texttt{ASteRA} struggles to constrain the spot filling factor for small spots at high latitudes (Cases 3, 12 and 21) and for small, low contrast spots (Cases 19, 20 and 21). For these scenarios the retrieved $F_{Spot}$ is highly degenerate as the comparatively low levels of contamination can be reproduced by a larger number of spot configurations. 

\begin{table}[htp]
\centering
\caption{Retrieved planetary and spot parameters for each of the 27 spot cases investigated. The ground truth (GT) planetary parameters are given for comparison.`Case 0' shows the parameters obtained when a retrieval using \texttt{ASteRA} is conducted on the uncontaminated spectrum as a frame of reference and to verify that \texttt{ASteRA} introduces no intrinsic bias. \smallskip}
\label{tab:retr_activity}
\begin{tabular}{llllllll}
\hline
Case No.     & $R_{P}$ ($R_{Jup}$) & $T$ (K)  & $log(H_{2}O)$ & \begin{tabular}[c]{@{}c@{}}Input \\ $T_{Spot}$ \\ (K)\end{tabular} & \begin{tabular}[c]{@{}c@{}}Retrieved \\ $T_{Spot}$ \\ (K)\end{tabular} & \begin{tabular}[c]{@{}c@{}}Input \\ $F_{Spot}$\end{tabular} & \begin{tabular}[c]{@{}c@{}}Retrieved \\ $F_{Spot}$\end{tabular} \\ \hline \hline
GT & 0.2730     & 400    & -3.00       & -               & -                   & -           & -               \\
0            & 0.2731$_{-0.0002}^{+0.0003}$     & 398.31$_{-5.62}^{+4.91}$ & -2.99$\pm$0.05    & N/A               & 4670.15$_{-19.95}^{+20.12}$                   & 0           & 0.391$_{-0.378}^{+0.400}$               \\
1            & 0.2729$_{-0.0005}^{+0.0004}$     & 399.74$_{-6.36}^{+5.74}$ & -3.01$\pm$0.05    & 3750            & 3918.02$_{-512.02}^{+442.34}$             & 0.010        & 0.018$_{-0.004}^{+0.012}$            \\
2            & 0.2732$_{-0.0005}^{+0.0004}$     & 395.98$_{-5.92}^{+5.58}$ & -2.99$\pm$0.05    & 3750            & 4017.22$_{-653.88}^{+490.05}$             & 0.009       & 0.015$_{-0.005}^{+0.030}$            \\
3            & 0.2736$\pm$0.0003     & 393.65$_{-6.00}^{+5.22}$ & -2.99$\pm$0.05    & 3750            & 4663.47$_{-585.54}^{+24.25}$             & 0.005       & 0.305$_{-0.299}^{+0.456}$            \\
4            & 0.2723$\pm$0.0005     & 403.97$_{-5.48}^{+5.68}$ & -3.05$\pm$0.05    & 3750            & 3779.57$_{-110.55}^{+112.28}$             & 0.040        & 0.059$_{-0.003}^{+0.004}$            \\
5            & 0.2728$_{-0.0005}^{+0.0004}$     & 398.95$_{-5.94}^{+5.62}$ & -3.01$\pm$0.05    & 3750            & 3760.82$_{-139.26}^{+140.09}$             & 0.035       & 0.045$_{-0.003}^{+0.004}$            \\
6            & 0.2734$\pm$0.0005     & 392.42$_{-6.34}^{+6.08}$ & -2.96$\pm$0.05    & 3750            & 3773.34$_{-384.27}^{+449.70}$             & 0.020        & 0.018$_{-0.004}^{+0.007}$            \\
7            & 0.2698$_{-0.0005}^{+0.0004}$     & 429.02$_{-5.45}^{+5.98}$ & -3.22$\pm$0.05    & 3750            & 3787.50$_{-59.60}^{+41.23}$             & 0.160        & 0.234$_{-0.006}^{+0.002}$            \\
8            & 0.2711$_{-0.0004}^{+0.0005}$     & 411.12$_{-4.94}^{+5.23}$ & -3.09$\pm$0.05    & 3750            & 3707.40$_{-39.19}^{+39.81}$             & 0.139        & 0.176$_{-0.002}^{+0.003}$           \\
9            & 0.2737$\pm$0.0004     & 385.34$_{-6.09}^{+5.98}$ & -2.88$\pm$0.05    & 3750            & 3689.36$_{-85.94}^{+77.58}$             & 0.080        & 0.070$\pm$0.003            \\
10           & 0.2731$_{-0.0005}^{+0.0004}$     & 396.84$_{-5.72}^{+6.09}$ & -3.01$\pm$0.05    & 4250            & 4301.79$_{-744.13}^{+308.63}$             & 0.010        & 0.013$_{-0.006}^{+0.031}$            \\
11           & 0.2733$_{-0.0005}^{+0.0004}$     & 396.01$_{-6.14}^{+5.58}$ & -3.01$\pm$0.05    & 4250            & 4534.88$_{-893.32}^{+139.69}$             & 0.009       & 0.018$_{-0.012}^{+0.491}$            \\
12           & 0.2733$\pm$0.0003     & 395.37$_{-5.64}^{+5.45}$ & -3.00$\pm$0.05    & 4250            & 4669.58$_{-48.39}^{+20.84}$             & 0.005       & 0.401$_{-0.387}^{+0.401}$            \\
13           & 0.2730$\pm$0.0004     & 398.71$_{-6.01}^{+5.72}$ & -3.02$\pm$0.05    & 4250            & 4517.02$_{-248.64}^{+97.03}$             & 0.040        & 0.100$_{-0.049}^{+0.092}$            \\
14           & 0.2731$\pm$0.0004     & 397.32$_{-5.79}^{+5.64}$ & -3.00$\pm$0.05    & 4250            & 4479.76$_{-277.08}^{+128.87}$             & 0.035       & 0.070$_{-0.035}^{+0.076}$            \\
15           & 0.2733$\pm$0.0004     & 394.98$_{-5.52}^{+5.48}$ & -2.98$\pm$0.05    & 4250            & 4299.29$_{-748.47}^{+288.83}$             & 0.020        & 0.015$_{-0.007}^{+0.031}$            \\
16           & 0.2716$_{-0.0005}^{+0.0006}$    & 413.66$_{-7.66}^{+6.10}$ & -3.13$_{-0.05}^{+0.06}$    & 4250            & 4311.13$_{-46.12}^{+72.60}$             & 0.160 & 0.221$_{-0.005}^{+0.044}$            \\
17           & 0.2724$_{-0.0005}^{+0.0007}$     & 404.22$_{-8.11}^{+6.42}$ & -3.05$\pm$0.06    & 4250            & 4310.73$_{-63.42}^{+97.73}$             & 0.139        & 0.171$_{-0.020}^{+0.037}$            \\
18           & 0.2736$_{-0.0005}^{+0.0004}$     & 391.16$_{-6.59}^{+6.20}$ & -2.95$\pm$0.05    & 4250            & 4341.04$_{-178.21}^{+235.08}$             & 0.080        & 0.073$_{-0.019}^{+0.146}$            \\
19           & 0.2733$_{-0.0004}^{+0.0003}$     & 396.28$_{-5.62}^{+4.78}$ & -3.01$_{-0.05}^{+0.04}$    & 4500            & 4657.20$_{-725.55}^{+29.61}$             & 0.010  & 0.145$_{-0.140}^{+0.592}$            \\
20           & 0.2733$\pm$0.0003     & 396.05$_{-5.85}^{+5.23}$ & -3.00$\pm$0.05    & 4500            & 4664.47$_{-452.27}^{+24.21}$             & 0.009       & 0.283$_{-0.278}^{+0.490}$            \\
21           & 0.2732$\pm$0.0003     & 396.10$_{-5.87}^{+5.37}$ & -2.99$\pm$0.05    & 4500            & 4670.68$_{-20.66}^{+19.06}$             & 0.005       & 0.404$_{-0.385}^{+0.402}$            \\
22           & 0.2729$\pm$0.0004     & 400.01$_{-6.01}^{+5.34}$ & -3.02$\pm$0.05    & 4500            & 4505.79$_{-276.72}^{+102.77}$             & 0.040        & 0.055$_{-0.028}^{+0.052}$            \\
23           & 0.2730$\pm$0.0004     & 398.58$_{-5.77}^{+5.35}$ & -3.01$\pm$0.05    & 4500            & 4456.58$_{-397.22}^{+149.41}$             & 0.035       & 0.033$_{-0.017}^{+0.047}$            \\
24           & 0.2734$_{-0.0005}^{+0.0003}$     & 395.19$_{-5.78}^{+5.29}$ & -3.00$\pm$0.05    & 4500            & 4632.19$_{-842.62}^{+49.95}$             & 0.020        & 0.035$_{-0.030}^{+0.603}$            \\
25           & 0.2724$\pm$0.0004     & 406.09$_{-5.21}^{+5.52}$ & -3.07$\pm$0.05    & 4500            & 4570.62$_{-105.43}^{+57.93}$             & 0.160        & 0.400$_{-0.192}^{+0.02}$            \\
26           & 0.2728$\pm$0.0004     & 401.39$_{-5.64}^{+5.21}$ & -3.03$\pm$0.05    & 4500            & 4562.99$_{-105.43}^{+57.93}$             & 0.139        & 0.302$_{-0.149}^{+0.027}$            \\
27           & 0.2733$_{-0.0004}^{+0.0003}$     & 398.53$_{-5.75}^{+4.93}$ & -2.98$\pm$0.05    & 4500            & 4501.57$_{-294.34}^{+111.42}$             & 0.080        & 0.065$_{-0.033}^{+0.066}$            \\ \hline
\end{tabular}
\end{table}

\subsection{Retrievals When the Spot Contamination is Neglected}
\label{results_nospot}

The retrieved parameters obtained for the 27 contaminated spectra when the spot parameters are not fit for are given in Table \ref{tab:retr_noactivity}. In comparison to the retrievals presented in Section \ref{results_spot}, the errors in the retrieved planetary parameters are far more significant, particularly for the highest activity cases. For the most contaminated case (Case 7), the decision to account for stellar activity, even with the simplified method used by \texttt{ASteRA}, can be the difference in retrieving an approximately solar level water abundance ($log(H_{2}O)$=-3.22$\pm 0.05$) or an incorrect subsolar water abundance ($log(H_{2}O)$=-5.32$\pm 0.06$) an underestimation equivalent to $>$ 2 orders of magnitude. The planetary temperature ($T_P$) is also significantly underestimated with the retrieved temperature of 275 K being 125 K cooler than the ground truth (400 K), which, in the context of a temperate planet represents a very substantial error ($>$ 30$\%$).

\begin{table}[htp]
\centering
\caption{Retrieved planetary parameters obtained for the same 27 cases as in Table~\ref{tab:retr_activity}) but without accounting for the presence of stellar contamination by fitting for the spot parameters. Comparison with the ground truth (GT) shows that in cases of severe activity (e.g. Cases 7, 8 and 9) the planetary parameters retrieved are highly inaccurate. \smallskip}
\label{tab:retr_noactivity}
\begin{tabular}{cccc}
\hline
Case No.     & $R_{P}$ ($R_{Jup}$) & $T$ (K)  & $log(H_{2}O)$ \\
\hline
\hline
GT & 0.2730    & 400    & -3.00    \\
0            & 0.2731$\pm$0.0002     & 398.40$_{-5.50}^{+5.08}$ & -3.00$\pm$0.05    \\
1            & 0.2746$_{-0.0002}^{+0.0003}$    & 390.68$_{-6.30}^{+5.63}$ & -3.09$\pm$0.05    \\
2            & 0.2743$_{-0.0002}^{+0.0003}$     & 389.63$_{-5.93}^{+6.06}$ & -3.05$_{-0.04}^{+0.05}$    \\
3            & 0.2736$_{-0.0002}^{+0.0003}$     & 392.96$_{-5.76}^{+5.61}$ & -3.00$_{-0.05}^{+0.04}$    \\
4            & 0.2792$\pm$0.0003     & 365.30$_{-7.10}^{+6.97}$ & -3.41$\pm$0.05    \\
5            & 0.2781$\pm$0.0003     & 368.64$_{-6.93}^{+6.72}$ & -3.27$\pm$0.05    \\
6            & 0.2753$\pm$0.0003     & 381.90$_{-6.23}^{+5.82}$ & -3.05$\pm$0.05    \\
7            & 0.2986$\pm$0.0001     & 274.94$_{-0.28}^{+0.18}$ & -5.32$\pm$0.06    \\
8            & 0.2944$\pm$0.0001     & 274.93$_{-0.41}^{+0.29}$ & -4.44$\pm$0.05    \\
9            & 0.2827$\pm$0.0003     & 330.52$_{-7.58}^{+7.90}$ & -3.28$\pm$0.07    \\
10           & 0.2738$_{-0.0002}^{+0.0003}$     & 393.36$_{-5.95}^{+5.17}$ & -3.04$\pm$0.05    \\
11           & 0.2737$\pm$0.0002     & 393.43$_{-5.80}^{+5.35}$ & -3.03$\pm$0.05    \\
12           & 0.2733$_{-0.0002}^{+0.0003}$     & 395.31$_{-6.02}^{+5.07}$ & -3.00$\pm$0.05    \\
13           & 0.2758$\pm$0.0003     & 385.69$_{-6.23}^{+6.47}$ & -3.24$\pm$0.05    \\
14           & 0.2753$\pm$0.0003     & 386.20$_{-6.26}^{+6.24}$ & -3.16$\pm$0.05    \\
15           & 0.2741$\pm$0.0001     & 390.73$_{-6.18}^{+5.56}$ & -3.04$_{-0.04}^{+0.05}$    \\
16           & 0.2833$_{-0.0003}^{+0.0004}$     & 374.04$_{-7.97}^{+7.45}$ & -4.16$_{-0.04}^{+0.05}$    \\
17           & 0.2819$\pm$0.0003     & 363.24$_{-7.01}^{+7.63}$ & -3.80$\pm$0.05    \\
18           & 0.2774$\pm$0.0003     & 370.91$_{-6.51}^{+6.50}$ & -3.21$\pm$0.05    \\
19           & 0.2735$\pm$0.0002     & 395.05$_{-5.87}^{+5.17}$ & -3.02$_{-0.04}^{+0.05}$    \\
20           & 0.2734$_{-0.0002}^{+0.0003}$     & 395.45$_{-5.87}^{+5.07}$ & -3.01$\pm$0.05    \\
21           & 0.2732$\pm$0.0002     & 396.01$_{-5.61}^{+5.06}$ & -2.99$_{-0.05}^{+0.04}$    \\
22           & 0.2745$_{-0.0002}^{+0.0003}$     & 395.50$_{-6.24}^{+5.09}$ & -3.13$\pm$0.04    \\
23           & 0.2742$\pm$0.0002     & 392.38$_{-5.78}^{+5.42}$ & -3.09$\pm$0.04    \\
24           & 0.2736$_{-0.0002}^{+0.0003}$     & 393.47$_{-5.78}^{+5.08}$ & -3.01$_{-0.05}^{+0.04}$    \\
25           & 0.2784$_{-0.0002}^{+0.0003}$     & 384.85$_{-6.03}^{+5.39}$ & -3.61$_{-0.04}^{+0.05}$    \\
26           & 0.2775$\pm$0.0003     & 382.73$_{-6.67}^{+7.00}$ & -3.44$\pm$0.05    \\
27           & 0.2752$\pm$0.0003     & 385.92$_{-6.34}^{+6.21}$ & -3.13$\pm$0.05    \\ \hline
\end{tabular}
\end{table}

\begin{figure}[htp]
    \centering
    \includegraphics[width=0.7\textwidth, height=0.43\textheight]{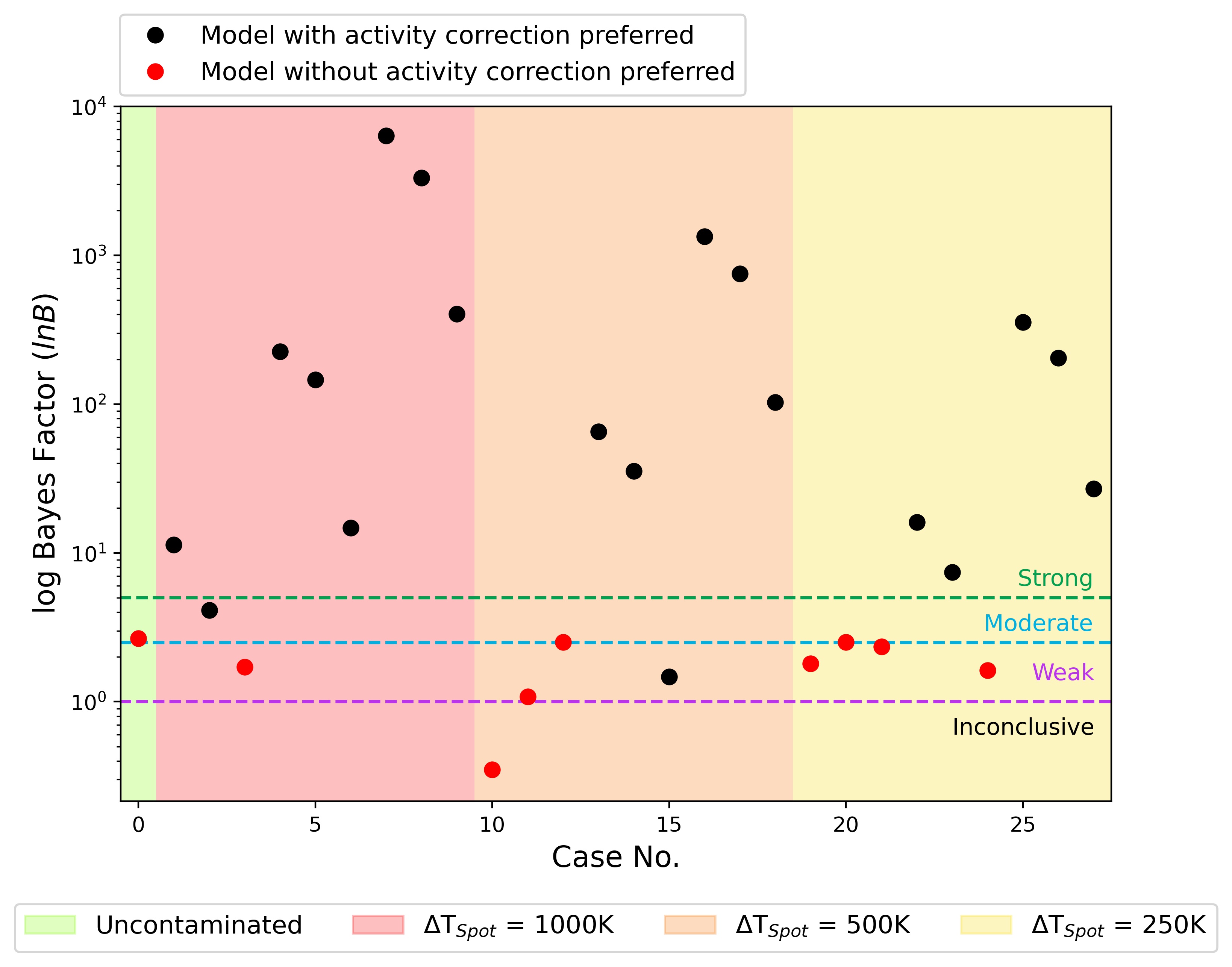}
    \caption{A graphical representation of the Bayes factors for the uncontaminated and 27 spot-contaminated cases explored. Black markers indicate that the Bayes factor is in favour of the ASteRA retrieval, where stellar activity has been accounted and corrected for. Red markers indicate a preference for the lower dimensionality retrieval in which the spot parameters are not fit for. The Jeffrey’s Scale \citep{trotta2008} is overplotted to show the strength of the model preference where a Bayes factor $\geq 1$ is indicative of weak preference, $\geq 2.5$ of moderate preference and $\geq 5$ a strong preference. A Bayes factor below 1 is deemed to be inconclusive.}
    \label{fig:Bayesfactors}
\end{figure}

Conducting two retrievals on the same contaminated spectrum allows us to determine which model is preferred by the data through the comparison of the Bayesian evidences. The Bayes factors ($lnB$) show a strong preference for the \texttt{ASteRA} retrieval for the majority of cases considered here (Fig~\ref{fig:Bayesfactors}). The cases in which only a weak preference for the corrected model is seen, or a preference for the retrieval in which contamination is neglected is indicated, correspond to those characterised by small, high latitude and low temperature contrast spots. As the contamination resulting from such spots is minimal, the model neglecting contamination is still able to perform reasonably well under conditions of low activity. In contrast, a penalty is applied to the model accounting for contamination when calculating the Bayesian evidence as a result of its higher dimensionality, explaining why the activity model is not conclusively preferred in the low activity regime despite a spot being present. For these low activity cases the planetary parameters are still accurately retrieved even when the presence of the spot is neglected entirely (Fig.~\ref{fig:retr_planet}). The Bayes factor shows the strongest preference for the retrieval without the activity correction for the uncontaminated case as one would expect. This reaffirms that no intrinsic bias is introduced by \texttt{ASteRA} and that \texttt{ASteRA} does not find evidence for stellar activity where there is none.

\section{Discussion}
\label{discussion}

\subsection{Understanding the Interaction Between Spot Contamination and the Limb Darkening Effect}
\label{LD-spot}

In order to adequately understand and interpret the retrieval results we must first consider how each of the three spot parameters will contribute to the contamination factor and therefore the observed spectrum. The size and the temperature contrast of the spot have very intuitive impacts on the contamination spectrum. The contamination factor increases as a function of both as expected. In comparison the effect of the spots latitude/position on the stellar disk is more subtle. It also affects only the shortest wavelengths with the largest variations seen at $\lambda \leq 1 \mu m$. The importance of the spot latitude in this study is twofold. Firstly, the projected filling factor (defined as the percentage of the 2D stellar disk that is covered by the spot) decreases as a function of latitude for a spot of a constant radius due to decreases in its 3D geometric projection onto the surface. This reduces the contamination introduced from the spatial coverage of the spot alone. Secondly, the position of the spot dictates how it will interact with the limb darkening effect as shown by the normalised intensity profiles (Fig.~\ref{fig:spotty_norm_int}) in Section \ref{StARPA}. We subsequently term this interaction as the limb darkening–spot interplay. The limb darkening–spot interplay is an important consideration as a central spot will remove a greater amount of flux compared to a spot (with an identical filling factor) located at the limb as the intensity maximum occurs at the disk centre.

\subsection{Accuracy of the Retrieved Planetary Parameters}
\label{retr_planet}

\begin{figure}[htp]
    \centering
    \includegraphics[width=1\textwidth, height=0.32\textheight]{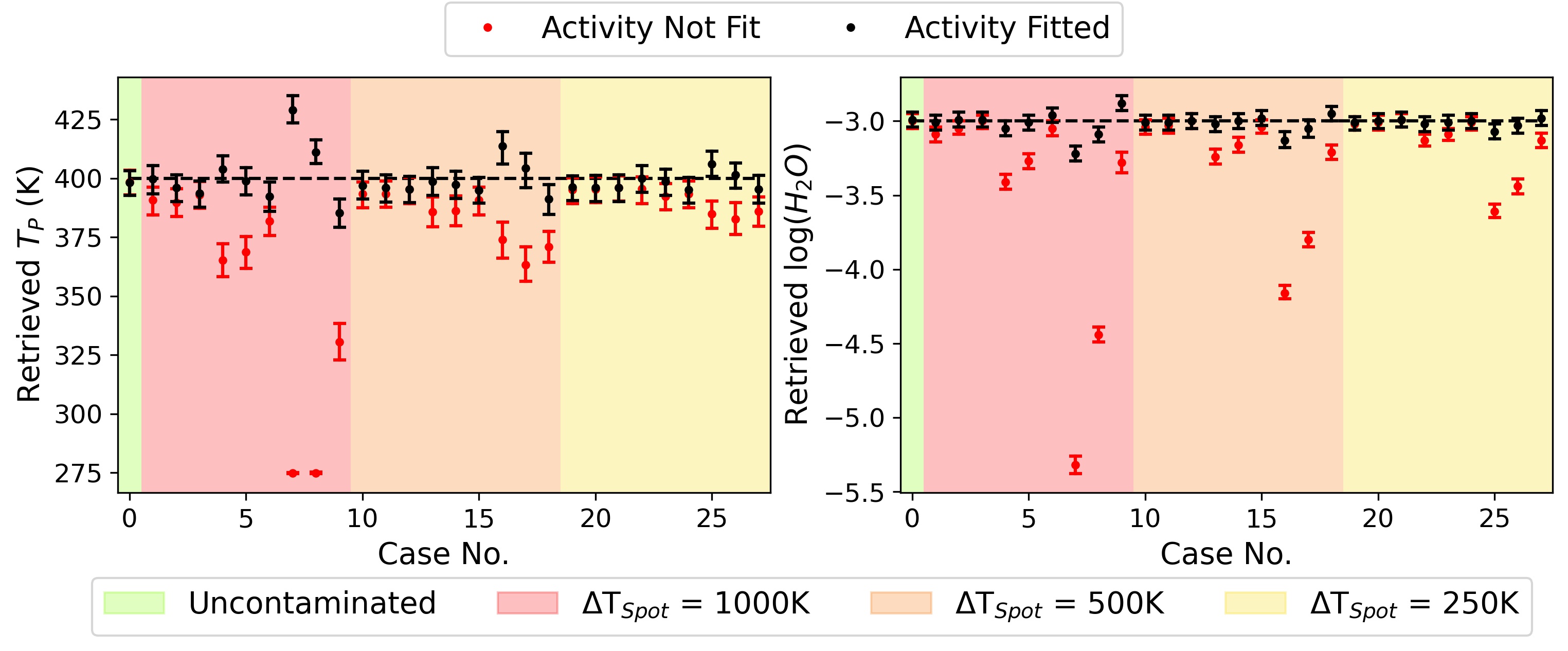}
    \caption{The retrieved planetary temperature ($T_{P}$) (left) and H$_{2}$O mixing ratio ($log(H_{2}O)$) (right) obtained for each spot case when the spot parameters are fit for (black data points) vs when activity is not accounted for (red data points). The plot background colours correspond to the temperature contrasts of the spot considered for each case. The ground truth for each parameter is indicated by the dashed black line.}
    \label{fig:retr_planet}
\end{figure}

In this section, we focus on how the use of the simplified spot model within \texttt{ASteRA} affects the retrieved planetary parameters (Fig.~\ref{fig:retr_planet}) and how these errors are observed to vary as a function of the spot parameters. As already stated in Section \ref{results_spot} the largest errors are seen for the most extreme activity cases. Fig.~\ref{fig:retr_planet} allows for the visual comparison of the values retrieved when the spot is fit for (black data points) as opposed to when it is not (red data points). This reiterates the large improvement in accuracy obtained through using even a simple activity correction method over no correction at all. In the worst case scenario not correcting for stellar activity leads to an underestimation of the water mixing ratio by over two orders of magnitude. Such a large error would significantly impact our understanding of the planet as a whole.
Another concerning result highlighted by Fig.~\ref{fig:retr_planet} is that, whilst the bias introduced by not correcting for stellar activity strongly affects the retrieval accuracy, it does not affect the retrieval precision, as evidenced by the small error bars. As such this high precision could be very misleading for real observations where a priori knowledge of the correct planetary parameters are not known. This is consistent with previous retrieval-oriented works e.g. \citet{iyer2020} and reaffirms that caution should be taken when analysing retrieval results where stellar contamination has not been included, even if the host star is thought to be less active.

\begin{figure}[htp]
    \centering
    \includegraphics[width=1\textwidth, height=0.35\textheight]{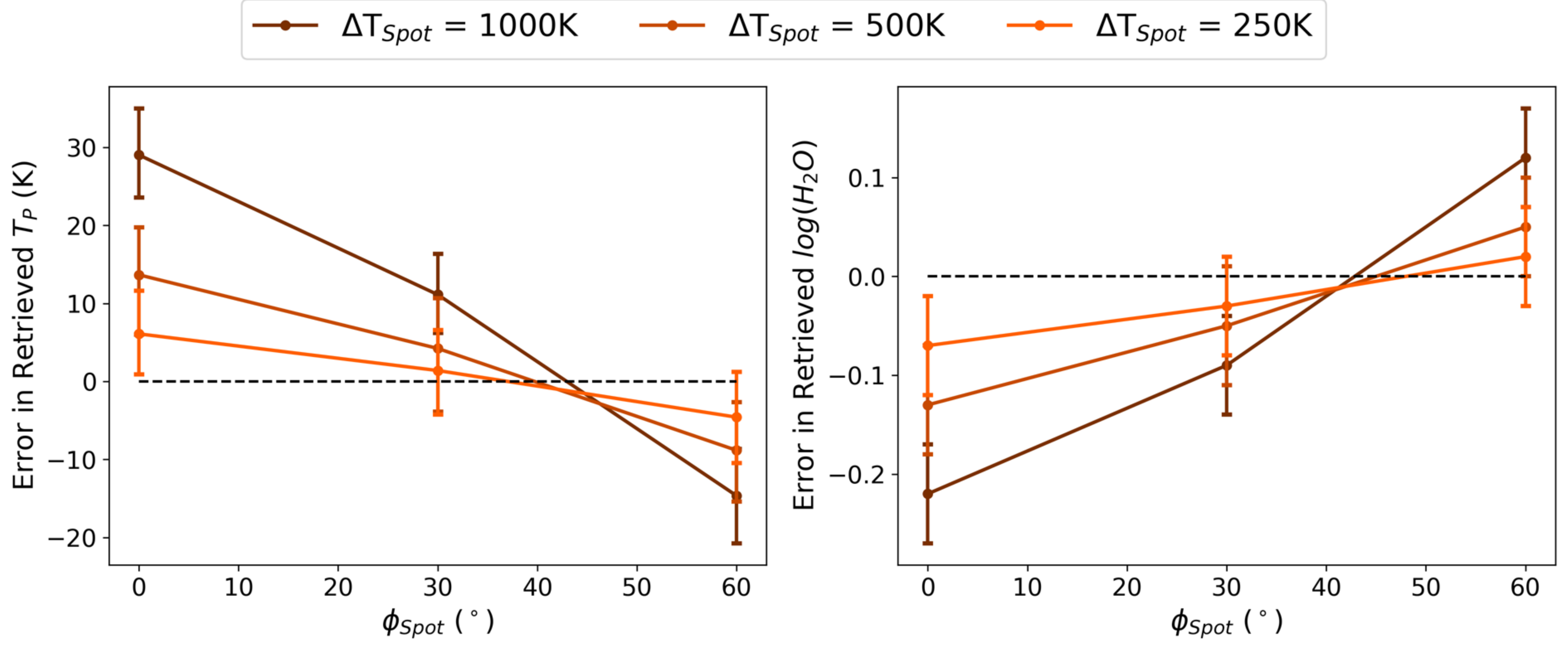}
    \caption{The error introduced in the retrieved planetary temperature $T_{P}$ (left) and atmospheric $H_{2}O$ mixing ratio $log(H_{2}O)$ (right) as a function of spot latitude. Only the cases with the largest spots (0.4 $R_{*}$) are plotted as these are the cases where some minor residual contamination remains after the correction. The effect of the spot temperature contrast on the observed trend is evident with the strongest trend seen for the highest contrast spots. There is a clear anticorrelation between $T_{P}$ and $log(H_{2}O)$.}
    \label{fig:retr_trends}
\end{figure}

The parameter that appears to be most influential in the highest activity cases is the spot latitude which is acting as a proxy for limb darkening (Fig.~\ref{fig:retr_trends}). It is intuitive that this spot parameter should have the greatest effect on the residual bias as \texttt{ASteRA} has no way of fitting for the spots position. A decreasing trend as a function of increasing $\phi_{Spot}$ is observed in the retrieved planetary temperature, with this being overestimated for the lower latitude spot cases (Cases 7 and 8) and subsequently underestimated at the highest latitude considered (Case 9). This underestimation can be attributed to the interplay of a high latitude spot and limb darkening. The reduced flux originating from the quiet photosphere at the limbs acts to reduce the observed spot contrast and thus there is also a reduction in the degree of contamination introduced. In contrast to this, the opposite trend is observed in the retrieved H$_{2}$O mixing ratio with this being underestimated at the two lower latitudes and overestimated at the highest. Similar trends in $T$ and $log(H_{2}O)$ are observed for the other cases considering large (0.4 $R_{*}$) spots (Cases 16, 17 and 18 and Cases 25, 26 and 27 respectively), however, the magnitude of the bias introduced is weaker due to the lower spot contrasts considered.

\newpage
\subsection{Accuracy of the Retrieved Stellar Parameters}

\begin{figure}[htp]
    \centering
    \includegraphics[width=1\textwidth, height=0.30\textheight]{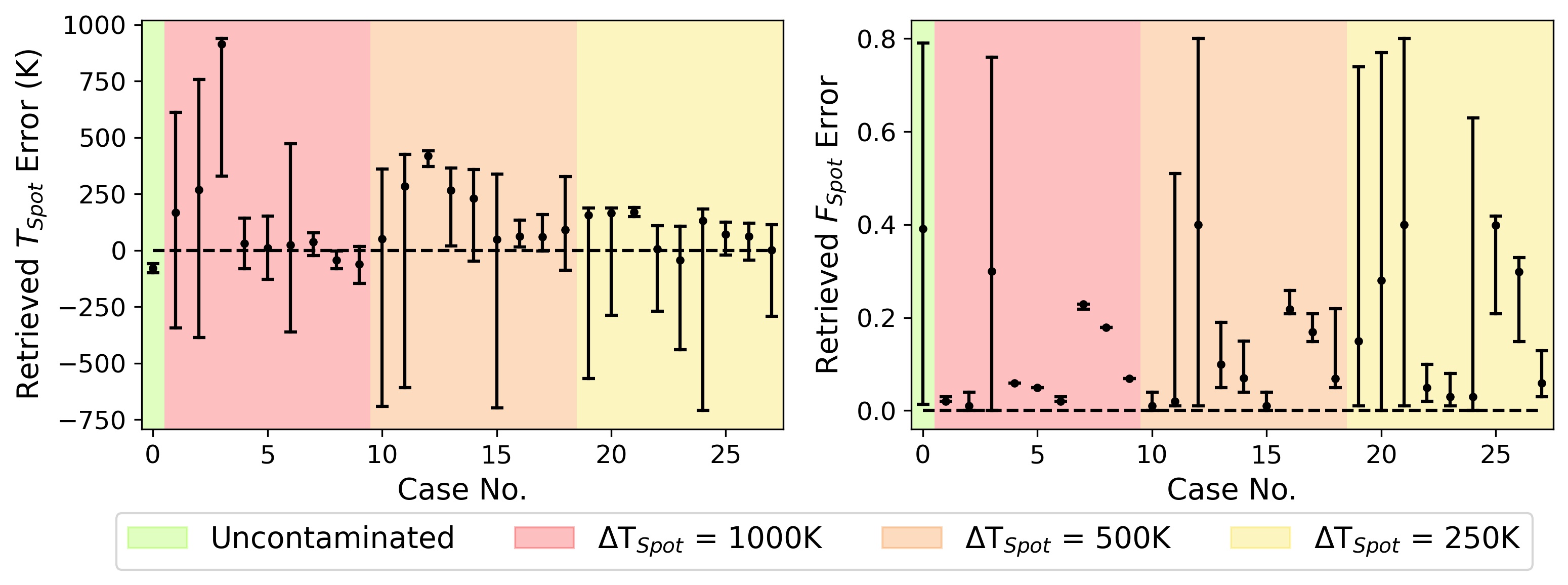}
    \caption{The error in the retrieved spot parameters $T_{Spot}$ and $F_{Spot}$ obtained for the uncontaminated and the 27 spot-contaminated cases. Figure elements are the same as in Fig.~~\ref{fig:retr_planet}. Note that $F_{Spot}$ is very poorly constrained for the uncontaminated case which is consistent with the absence of stellar activity. In contrast, $T_{Spot}$} 
    \label{fig:retr_stellar}
\end{figure}

\begin{figure}[htp]
    \centering
    \includegraphics[width=0.95\textwidth, height=0.47\textheight]{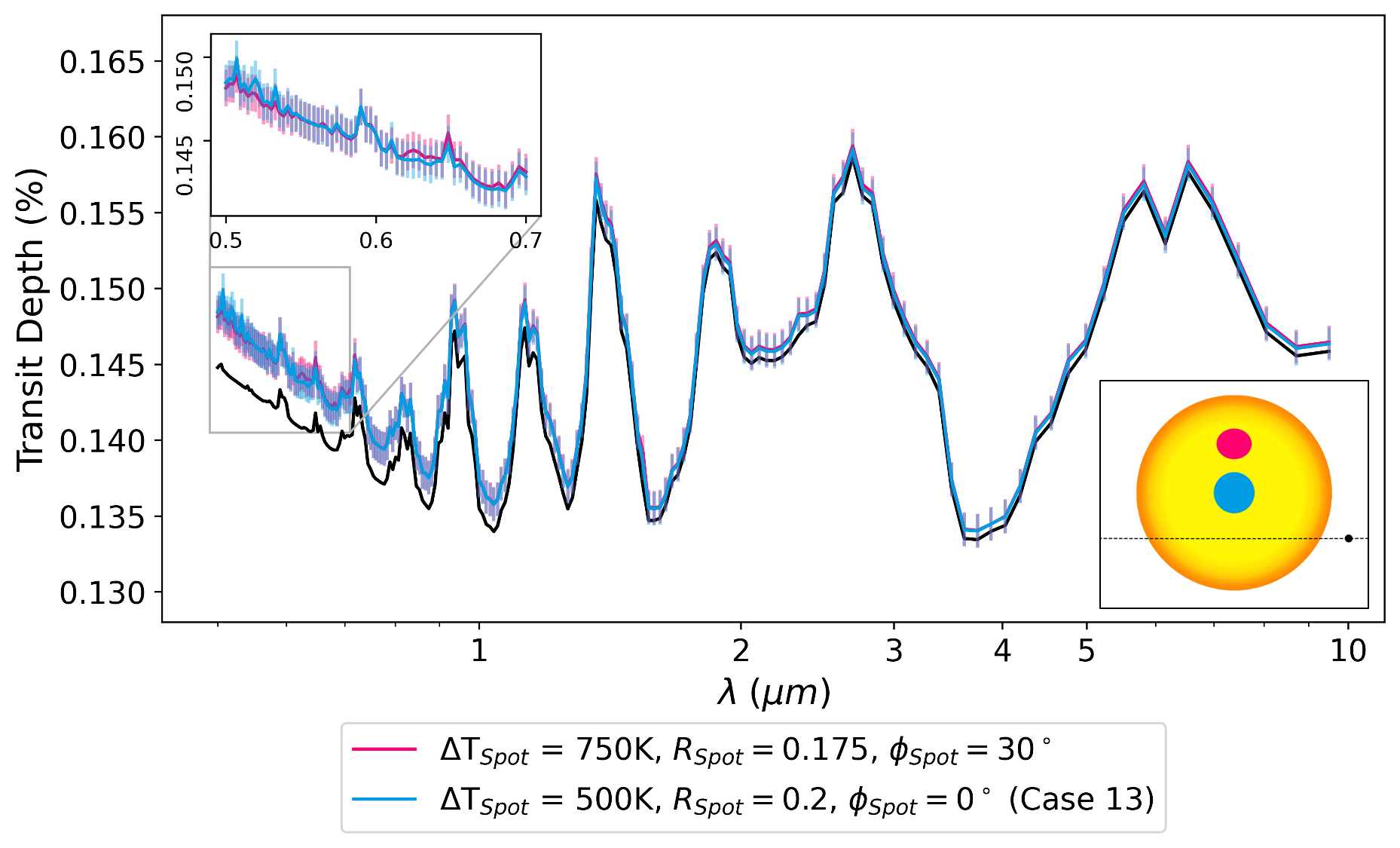}
    \caption{An example of how spot parameters can be degenerate at low resolution, particularly in cases of moderate activity. The black line depicts the uncontaminated transmission spectrum. In blue is the contaminated spectrum obtained with a slightly larger ($R_{Spot} = 0.2 R_{*}$), central ($\phi_{Spot} = 0\degree$) spot with a lower temperature contrast ($\Delta T_{Spot} = 500$ K) which is equivalent to Case 13 in our retrieval grid. In contrast to this the pink contaminated spectrum results from a smaller ($R_{Spot} = 0.175 R_{*}$), higher latitude ($\phi_{Spot} =30\degree$), higher contrast ($\Delta T_{Spot} = 750$ K) spot. The two spot-contaminated spectra are undifferentiable beneath the noise at all wavelengths. As such, a retrieval would not be able to confidently differentiate between these two solutions. \textit{Inset (top left)} – A close up view of the two spot contaminated spectra at 0.5-0.7$\mu m$ where they diverge the most. The differences between them are still easily lost beneath the very optimistic noise estimate considered here (10ppm). \textit{Inset (bottom right)} – A visual depiction of the two degenerate spots.}
    \label{fig:degeneracy}
\end{figure}

\texttt{ASteRA} is less successful in recovering the spot parameters (Fig.~\ref{fig:retr_stellar}). This is likely due to a combination of not accounting for limb darkening and also because many combinations of the three spot parameters are degenerate at low resolution, particularly if the spot in question has a low to moderate temperature contrast. An example of how different spot parameter combinations can result in degenerate solutions is given in Fig.~\ref{fig:degeneracy} for clarity. The largest errors and uncertainties in the retrieved $T_{Spot}$ are seen for the smallest spots considered, especially when present at high latitudes. For larger spots, $T_{Spot}$ is generally reasonably well constrained due to the larger contamination effects they introduce. Large errors and uncertainties are also seen in the retrieved $F_{Spot}$ values in several cases. Large error bars point to substantial degeneracy in the small spot cases, whereas, in the case of the largest spots $F_{Spot}$ is often constrained to a higher precision but significantly overestimated. The results of these retrievals indicate that we should be more cautious with retrieved stellar parameters, as the degeneracies between them mean that they are constrained less confidently by the retrieval. Simultaneous, external observations e.g. at different bandpasses could help to break some of these degeneracies.

\subsection{The Effect of Limb Darkening}
\label{disc_LD}

\begin{figure}[htp]
    \centering
    \includegraphics[width=0.56\textwidth, height=0.8\textheight]{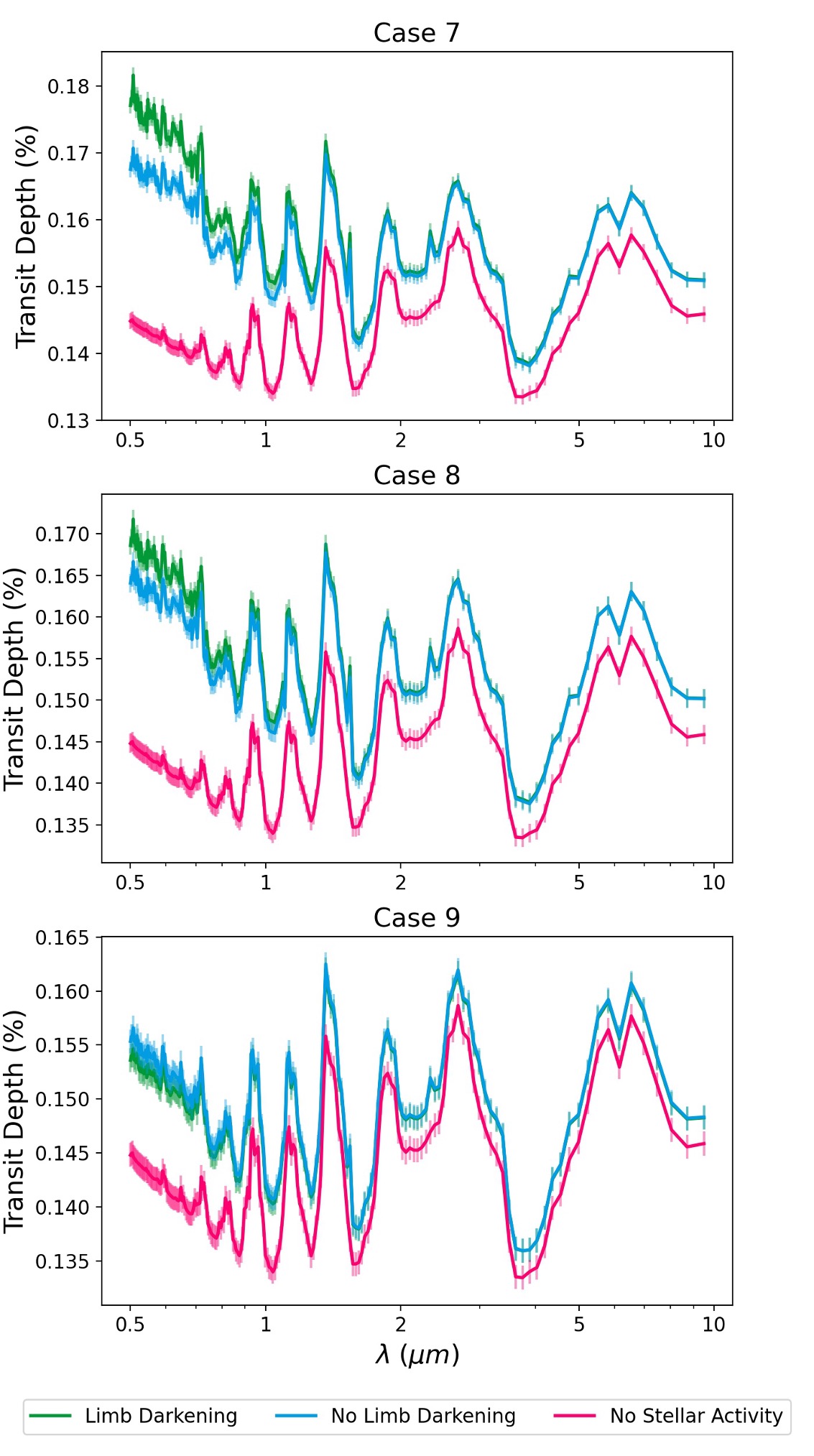}
    \caption{Forward model transmission spectra highlighting the effect of accounting for the contribution of limb darkening when considering the spot contamination or not for the most severely contaminated cases (Cases 7, 8 and 9). The uncontaminated spectrum assuming a quiet host star is shown in pink. The green spectrum is the spot-contaminated spectrum produced when the contribution of the limb darkening effect to the overall contamination is considered, the blue spectrum, in contrast, is the contaminated spectrum obtained when limb darkening is not accounted for. The error bars are equivalent to 10 ppm.}
    \label{fig:limbdarkening}
\end{figure}

In order to attribute the biases seen in the planetary parameters to not accounting for the effect of limb darkening, the worst-case scenarios (Cases 7, 8 and 9) were rerun. For each case the contaminated spectrum was regenerated with the limb darkening coefficients set to zero (Fig.~\ref{fig:limbdarkening}) and a further retrieval conducted. When noise is taken into account the effect of including vs. excluding the limb darkening effect is really only distinguishable at the shortest wavelengths considered ($\lambda <$ 1 $\mu m$), even for these worst-case scenarios. Ariel in particular will only be able to access this wavelength region through three photometric bands \citep{tinetti2021}, as such, extracting as much information as possible about the activity of the host star from these three data points will be of utmost importance. The retrieved parameters when the limb darkening effect is removed are given in Table \ref{tab:LD_789}. With the exception of $T_{Spot}$, which was already well constrained, all other parameters are retrieved more accurately, including $F_{Spot}$ now being constrained correctly. As the input observations are inherently different due to the removal of the limb darkening effect, unfortunately we cannot use the Bayes Factor as a means of model comparison here as was done in previous sections (Fig.~\ref{fig:Bayesfactors}). However, the fact that both the planetary and stellar parameters are retrieved more accurately when the limb darkening contribution is removed from the input observation provides compelling evidence in itself that the residual bias seen originates from the limb darkening-spot interplay. For real observations ignoring the effects of limb darkening within activity correction frameworks will unfortunately not be a viable solution as this phenomenon will always be present. As such, we believe that as a community there is a need for us to push towards developing and using stellar activity models in which the limb darkening-spot interplay is accounted for, particularly when dealing with highly active host stars.

\begin{table}[htp]
\caption{The retrieved spot and planetary parameters obtained for Cases 7,8 and 9 when the effect of limb darkening is no longer present in the input spectra i.e. the blue spectra from Fig.~\ref{fig:limbdarkening}. The ground truth (GT) spot and planetary parameters are given for comparison. $F_{Spot}$ varies on a case by case basis, as such, the Input $F_{Spot}$ denotes the ground truth value for each case. \smallskip}
\label{tab:LD_789}
\begin{tabular}{ccccccc}
Case No. & $T_{Spot}$ & Input $F_{Spot}$ & Retrieved $F_{Spot}$ & $R_{P}$ ($R_{Jup}$) & $T$ & $log(H_{2}O)$ \\ \hline 
\hline
GT & 3750 & - & - &  0.273 & 400 & -3 \\
7  & 3700.81$_{-33.72}^{+34.68}$ & 0.160 & 0.166$\pm$0.002 & 0.2728$\pm$0.0004 & 392.47$_{-6.06}^{+5.69}$ & -2.91$\pm$0.05 \\
8  & 3704.40$_{-38.06}^{+33.81}$ & 0.139 & 0.144$\pm$0.002 & 0.2728$\pm$0.0004 & 393.19$_{-5.51}^{+6.37}$ & -2.92$\pm$0.05 \\
9  & 3715.14$_{-58.94}^{+82.23}$ & 0.080 & 0.084$\pm$0.003 & 0.2730$\pm$0.0004 & 395.72$_{-6.15}^{+5.58}$ & -2.95$\pm$0.05 \\ \hline
\end{tabular}

\end{table}

The posterior distributions retrieved for the two input scenarios, spot-contaminated with the limb darkening contribution and spot-contaminated with the limb darkening contribution removed, are given in Appendix \ref{appendix}. The greatest deviation between the two posteriors is seen in the case of a central spot (Fig. \ref{fig:case7_posteriors}). This is intuitive as the spot masks the disk centre where a greater proportion of the stellar flux originates from and as such, results in the largest residual contamination due to not encompassing the limb darkening-spot interplay. As the spot is modelled at progressively higher latitudes, less residual contamination remains and the posteriors for the limb darkened, contaminated spectra (Fig. \ref{fig:case8_posteriors} and Fig. \ref{fig:case9_posteriors}) converge towards the correct values.

\clearpage

\subsection{Preliminary Investigations into Multiple Spot Cases}
\label{multispot}

\begin{figure}[htp]
    \centering
    \includegraphics[width=1\textwidth, height=0.32\textheight]{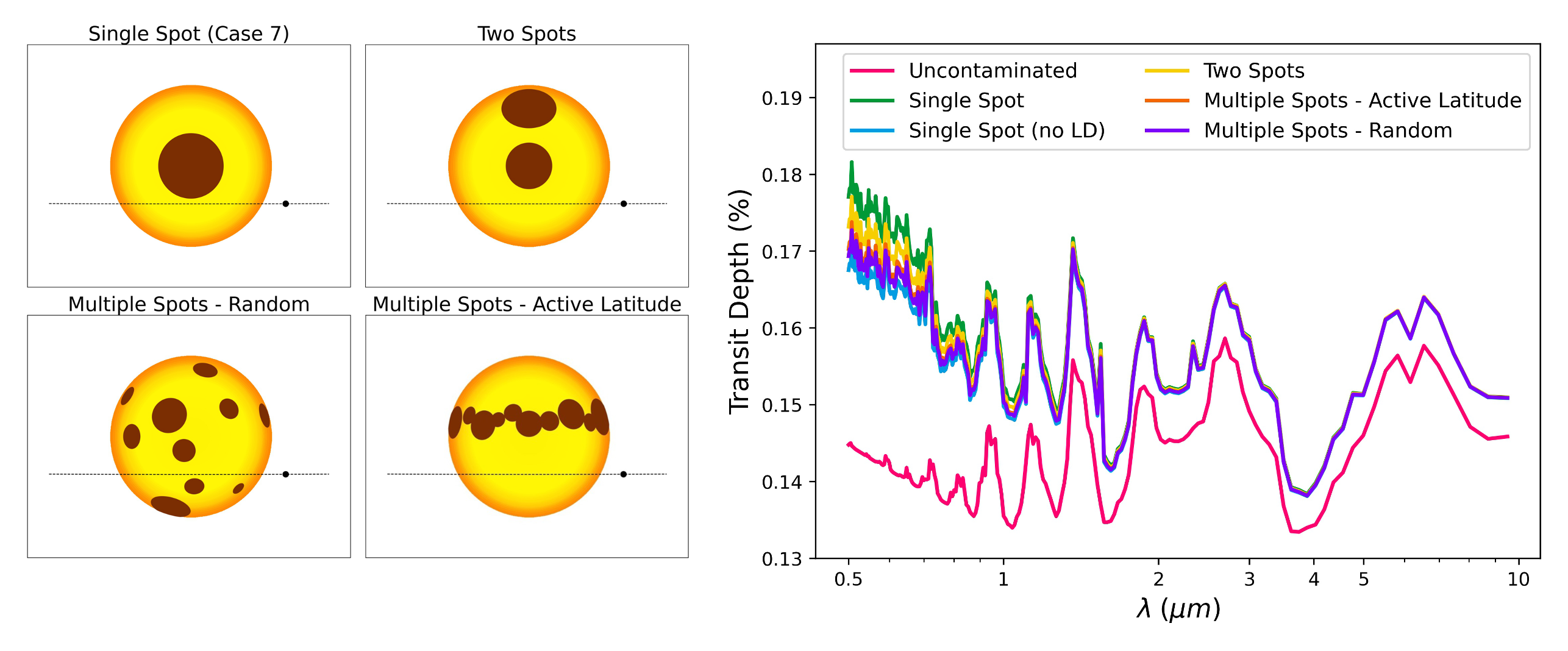}
    \caption{\textit{Left} – Visual representations of the single spot Case 7 and the three multiple spot cases investigated: a two spot case, a multiple spot case with a random spot configuration and a multiple spot case where the spots occupy a preferred active latitude centred around $\phi = 15\degree$. All spots have the same temperature contrast ($\Delta T_{Spot}=1000K$) and their combined filling factors are kept constant at $F_{Spot}=0.16$. \textit{Right} – The resulting contaminated spectra for the two spot (yellow), active latitude (orange) and random configuration (purple) multiple spot cases compared to the uncontaminated spectrum (pink) and the single spot of Case 7 with (green) and without (blue) the contribution of the limb darkening effect. The 10ppm error bars have been removed here for clarity.}
    \label{fig:multispots}
\end{figure}

In previous sections we have focused only on isolated, single spots in order to assess the interactions of the spot parameters and how they influence the observed, contaminated spectrum. In order to better understand the limb darkening-spot interplay (Section \ref{LD-spot}) and the bias it introduces in retrievals (Section \ref{disc_LD}) we conducted three preliminary multiple spot cases using the same methodology to explore how this interplay may differ when more than one spot is present. For comparability the spots have the same temperature ($T_{Spot}$ = 3750 K) and total filling factor ($F_{Spot}$ = 0.16) as the single spot Case 7. The multiple spot cases we consider are a two spot case and two cases characterised by 10 smaller spots. These 10 spots are arranged in a random configuration on the stellar disk in the first scenario, and occupy a preferred active latitude centred around $\phi = 15\degree N$ in the second scenario. The only strict requirement for these cases is that all of the spots remain unocculted. The resultant contaminated spectra are shown in Fig.~\ref{fig:multispots} alongside the uncontaminated spectrum and the Case 7 contamination spectrum, with and without the contribution of the limb darkening effect, as in Fig.\ref{fig:limbdarkening}. These spectra show that, as the number of spots increases and they are distributed over a larger range of $\mu$ values (i.e. located at different distances from the disk centre) 
the limb darkening-spot interplay and its contribution to the contamination spectrum are minimised. A table of the retrieved parameters (Table~\ref{tab:multispot}), alongside the posterior distributions (Figs. \ref{fig:twospotposterior}, \ref{fig:multirandomspotposterior} and \ref{fig:activelatposterior}) are given in Appendix \ref{appendixB}.

These additional multiple spot experiments highlight that the large, high contrast single spots considered as the main focus of this study likely do represent the worst case scenario when considering the additional complication arising from the limb darkening effect. Observationally, the regimes in which the limb darkening-spot interplay will really start to matter will therefore be those in which the system geometries most closely replicate those in cases 7 and 8. Large high latitude and polar spots have been proposed to exist on several stars \citep[e.g.][]{jarvinen2018, almenara2022, strassmeier2023} although any biases that these could introduce will be lessened if the stars rotation axis is not significantly inclined with respect to the line of sight as the spot will appear at the limb. The worst-case scenario would occur for a system where the host star both possesses a polar spot and is inclined relative to the transiting planet and observer. In such a scenario the polar spot could manifest as a large central spot. Examples such as the HAT-P-11 system which consists of an active K4-dwarf hosting two highly misaligned planets \citep[e.g.][]{sanchisojeda2011, morris2017, yee2018} and the almost pole-on, solar-type star $\tau$ Ceti which is frequently included in target lists for exoplanet searches \citep{korolik2023} both indicate that observing a system with such geometry in the future cannot be ruled out.

Our recommendation for dealing with this worst-case, high activity regime, where it will be necessary to include the limb darkening-spot interplay in order to fully negate the bias introduced by stellar contamination, is to improve our retrieval stellar models so that they are capable of also parameterising the position of spots on the stellar disk. This is the focus of ongoing work, however, as with increasing the dimentionality of any model, we cannot ignore the risk of injecting intrinsic bias if the additional complexity is not physically motivated. For real observations there will also be the added challenge of not having any a priori knowledge of spot positions as the stellar disks are not spatially resolved. This work has shown that at first order using the simpler, two parameter spot prescription as is done with \texttt{ASteRA} is sufficient to remove the majority of the bias and enable a good understanding of the planets atmospheric properties. A push towards a better understanding of the limb darkening effect, whether in the presence of stellar activity or not, will also be highly beneficial. This will be especially important for later type stars where offsets due to the choice of limb darkening treatment appear to be inevitable \citep{patel2022}.

\subsection{Preliminary Investigation into Spots Displaying a Radial Temperature Variation - A Separation into Umbra and Penumbra}
\label{multitempspot}

All of the retrievals presented in the previous sections have been conducted assuming a single $T_{Spot}$. In order to explore the validity of this assumption, we conducted preliminary studies into modelling the contamination introduced by spots with radial temperature variations. This section examines the resulting impact on retrieval performance. We assume the same spot geometry as was used in the highest contamination, worst case scenario (Case 7). The spot is now separated into a cooler central umbra characterised by a temperature $T_{Umbra}$ = 3750 K, and a surrounding penumbra characterised by a temperature $T_{Pen}$ = 4250 K. This separation into two regions of distinct temperatures is consistent with observed sunspots and with the 3D MHD models for later type stars produced by \citet{panja2020}. We explore two different cases in which the total spot filling factor ($F_{Spot}$) is kept constant but the relative fractions of the umbra ($F_{Umbra}$) and penumbra ($F_{Pen}$) are varied. Table \ref{tab:tempvarinputs} shows the input parameters for these two cases. In doing this, we explore the retrieval response to contamination effects resulting from a spot that is characterised by two different temperatures/SEDs when the retrieval model is only capable of fitting this contamination using a single $T_{Spot}$ value. 

As expected, this further discrepancy between the forward model and the retrieval model introduces an additional source of error to the retrieved planetary parameters. The retrieved parameters are given in Table \ref{tab:tempvarretrieved} alongside the equivalent single $T_{Spot}$ case (Case 7) for comparison. The posterior distributions for the two spot temperature variation retrievals are given in Appendix \ref{appendixC}. The retrieved spot parameters give insight into how the retrieval has attempted to correct for the dual temperature spot. In both cases, the retrieval is best able to fit for the contamination with a moderate $T_{Spot}$ close to $T_{Pen}$, and an overestimated $F_{Spot}$. The overestimate in the spot filling factor accommodates the increased contamination due to the higher contrast umbra. The posteriors show an increasing correlation and degeneracy between $T_{Spot}$ and $F_{Spot}$ compared to the single $T_{Spot}$ cases. The introduction of radial temperature variation slightly enhances the biases in the retrieved planetary parameters that were observed for the highest activity cases. Intuitively, the greater effect is seen when considering a spot with a higher umbra-penumbra ratio, resulting in an overestimate in the retrieved planetary temperature of $\sim$ 47 K and an underestimate in the water mixing ratio of 0.39 magnitudes. Although this error is non-negligible, importantly it is not limiting. Retrievals conducted in the presence of such contamination would still be useful and informative.

\section{Conclusion}
\label{conclusion}
The main objective of this paper is to determine how complex our stellar activity models should be in order to remove biases so that we can characterise the transiting planet as accurately and efficiently as possible. At the same time, we want to ensure that our retrieval analysis remains reliable. As such, it is important that we are not introducing unjustified complexity which may inject a bias intrinsically. We make use of a grid of 27 spot-contaminated stellar disks created with the more complex forward stellar model (\texttt{StARPA}), in which the interplay of the spot and limb darkening is accounted for, and conduct retrievals with a simpler model (\texttt{ASteRA}) that neglects this in order to explore under which conditions this additional complexity is necessary. We find that \texttt{ASteRA} performs very well in cases of weak to moderate stellar contamination, constraining the planetary parameters to a high degree of accuracy. Importantly, in all cases, \texttt{ASteRA} performs far better than no correction attempt at all, consistent with the findings of previous studies. The need for an activity correction is especially demonstrated by the retrieved H$_{2}$O mixing ratios which are underestimated by over two orders of magnitude in the worst-case scenario when no correction is applied. \citet{iyer2020} find that stellar contamination must be accounted for when spot filling factors exceed 1$\%$ in order to avoid biasing the retrieved planetary parameters. Our results substantiate this, however, we highlight that the spot temperature contrast is also important to consider alongside the filling factor. The Bayes Factor is still moderately ($lnB$= 4.12) to strongly ($lnB$=11.33) in favour of the model containing the activity correction in the cases of a small (0.1$R_{*}$), high contrast ($\Delta T_{Spot}$=1000K) spot at latitudes of 30$\degree$ and 0$\degree$ respectively (Cases 1 and 2), despite both of these spots having filling factors $\leq$ 1$\%$. The spot latitude also contributes but to a lesser degree than the temperature contrast in the small spot regime. Importantly, there is no evidence for a hard boundary or ‘on-off switch’ for where stellar contamination should be considered. It will therefore be beneficial to apply an activity correction to safeguard against bias even in the lower activity regimes where contamination is less dominant. 

Degeneracies between spot parameters at low resolution means that the retrievals tend to be less successful for accurately characterising the host star, particularly in regimes of low to moderate activity. Nevertheless, from a planetary perspective it is an excellent tool for accounting for potential contamination bias in the fundamental planetary properties $R_{P}$, $T_{P}$ and $log(H_{2}O)$. We believe that this is how it should be viewed and utilised by the community moving forward, albeit with the one caveat that it cannot remove any bias that is introduced as a result of the inaccurate characterisation of fundamental stellar parameters.

For the highest activity cases considered here, the bias introduced by stellar contamination cannot be fully corrected for by \texttt{ASteRA} due to the limb darkening-spot interplay being neglected. As such, in these scenarios, a small amount of residual bias remains resulting in a slight loss of accuracy in the retrieved planetary parameters. This subtle loss of accuracy would not be easily identified without a priori knowledge of the activity level of the star. For this reason, it may be beneficial to consider other stellar activity mitigation processes in parallel if these are available e.g. utilising the out-of-transit observations or continuous photometry, in order to better interpret the retrieval results in the context of the host star. We show that for multiple spot cases the bias introduced by the limb darkening-spot interplay is minimised under the assumption that all spots have the same temperature. As such, single large spots present the worst case scenario for real observations. Further bias is also introduced if these large spots have separated into umbral and penumbral regions characterised by different distinct temperatures, making this something we should progressively start to consider as a community.

Although the results of this study are based on idealised spectra, the spot cases investigated in this analysis will provide a good baseline from which to fully explore the impact of stellar activity on both JWST and Ariel observations as our simulations cover a similar wavelength range and spectral resolution to what is/will be obtainable with these observatories. In future work we intend to conduct similar investigations using more realistic models and observations. With this in mind, from a stellar perspective we aim to extend the flexibility of the \texttt{ASteRA} plugin to incorporate the interplay of spots and limb darkening, as this study has shown that there are scenarios where this cannot reasonably be neglected. This will, however, need to be done with caution to avoid unknowingly injecting bias. We also aim to extend our retrievals to investigate other spectral types, in particular M dwarfs which may be more complicated, especially as their spectra can contain molecular lines that could be incorrectly attributed to the exoplanet atmosphere. Other exciting lines of investigation that naturally follow on from this work include exploring more complex manifestations of stellar activity for example; occulted spots and the presence of both spots and faculae. With respect to the exoplanetary atmosphere, we intend to extend this analysis to more complex, realistic atmospheric compositions. A particular emphasis will be put on physical processes responsible for producing features in the optical regime, where the stellar contamination is most pronounced, such as the presence of clouds and hazes and other opacity sources e.g. absorption by the alkali metals Na and K. Finally, we intend to transition from using idealised spectra to simulated instrument observations to explore the effects of more realistic noise and a more restricted wavelength coverage in the optical before eventually using this framework to achieve our ultimate goal of accurately analysing real observations.\\

\section*{Acknowledgements}
The authors would like to thank the anonymous referee for their insightful comments. AT would like to thank Dr Kai Hou Yip for their insightful discussions. The work presented in this paper was partially supported by UKSA, grant ST/X002616/1. The authors also acknowledge the support of the ARIEL ASI-INAF agreement n.2021-5-HH.0. G. M. has received funding from the European Union's Horizon 2020 research and innovation programme under the Marie Sk\l{}odowska-Curie grant agreement No. 895525, and from the Ariel Postdoctoral Fellowship program of the Swedish National Space Agency (SNSA).
This work would not have been possible without the use of publicly available data and open source software, as such the authors also wish to reiterate their acknowledgement of the use of TauREx3 \citep{alrefaie2021} \url{https://github.com/ucl-exoplanets/TauREx3_public}, ExoTETHyS \citep{morello2020a,morello2020b,morello2021} \url{https://github.com/ucl-exoplanets/ExoTETHyS}, pylightcurve \citep{tsiaras2016} \url{https://github.com/ucl-exoplanets/pylightcurve}, the PHOENIX BT-Settl Library \citep{allard2012, baraffe2015}, MultiNest \citep{feroz2009} and PyMultiNest \citep{buchner2014}.

{}

\appendix

\section{Posterior Distributions Retrieved with ASteRA}
\label{appendix}

\begin{figure}[htp]
    \centering
    \includegraphics[width=0.95\textwidth, height=0.77\textheight]{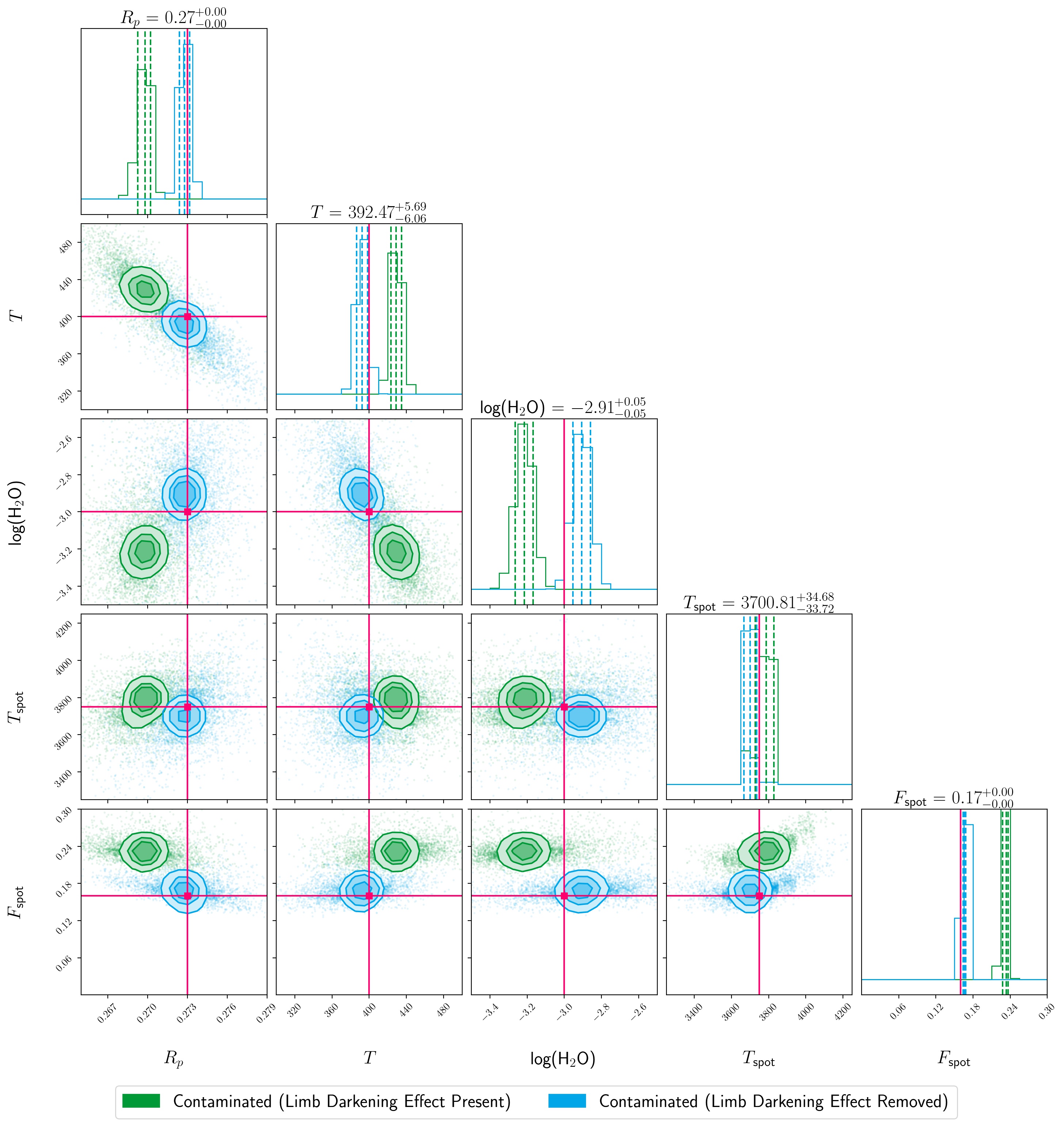}
    \caption{Retrieved posterior distributions for spot Case 7 when retrievals were conducted on a spot-contaminated spectrum including contributions from the interplay of the spot and the limb darkening effect (green) and for a spot-contaminated spectrum when the limb darkening contribution is removed i.e. the limb darkening coefficients are set to zero (blue). The retrieved values given for each parameter above each column correspond to the blue posteriors. The ground truth values for each parameter are indicated by the pink lines.}
    \label{fig:case7_posteriors}
\end{figure}

\begin{figure}[htp]
    \centering
    \includegraphics[width=0.95\textwidth, height=0.77\textheight]{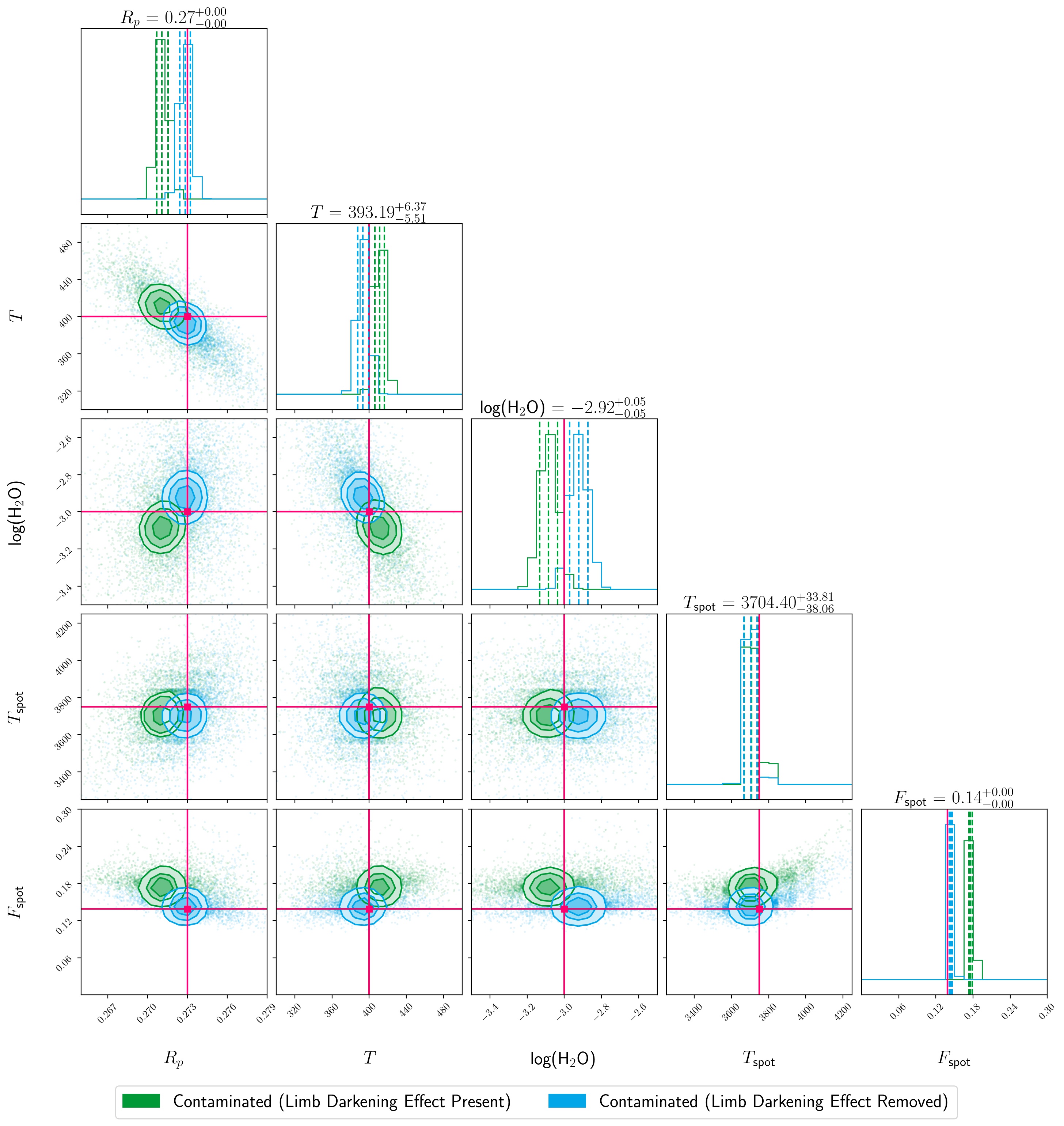}
    \caption{Retrieved posterior distributions for spot Case 8. Figure elements are the same as those in Fig. \ref{fig:case7_posteriors}}
    \label{fig:case8_posteriors}
\end{figure}

\begin{figure}[htp]
    \centering
    \includegraphics[width=0.95\textwidth, height=0.77\textheight]{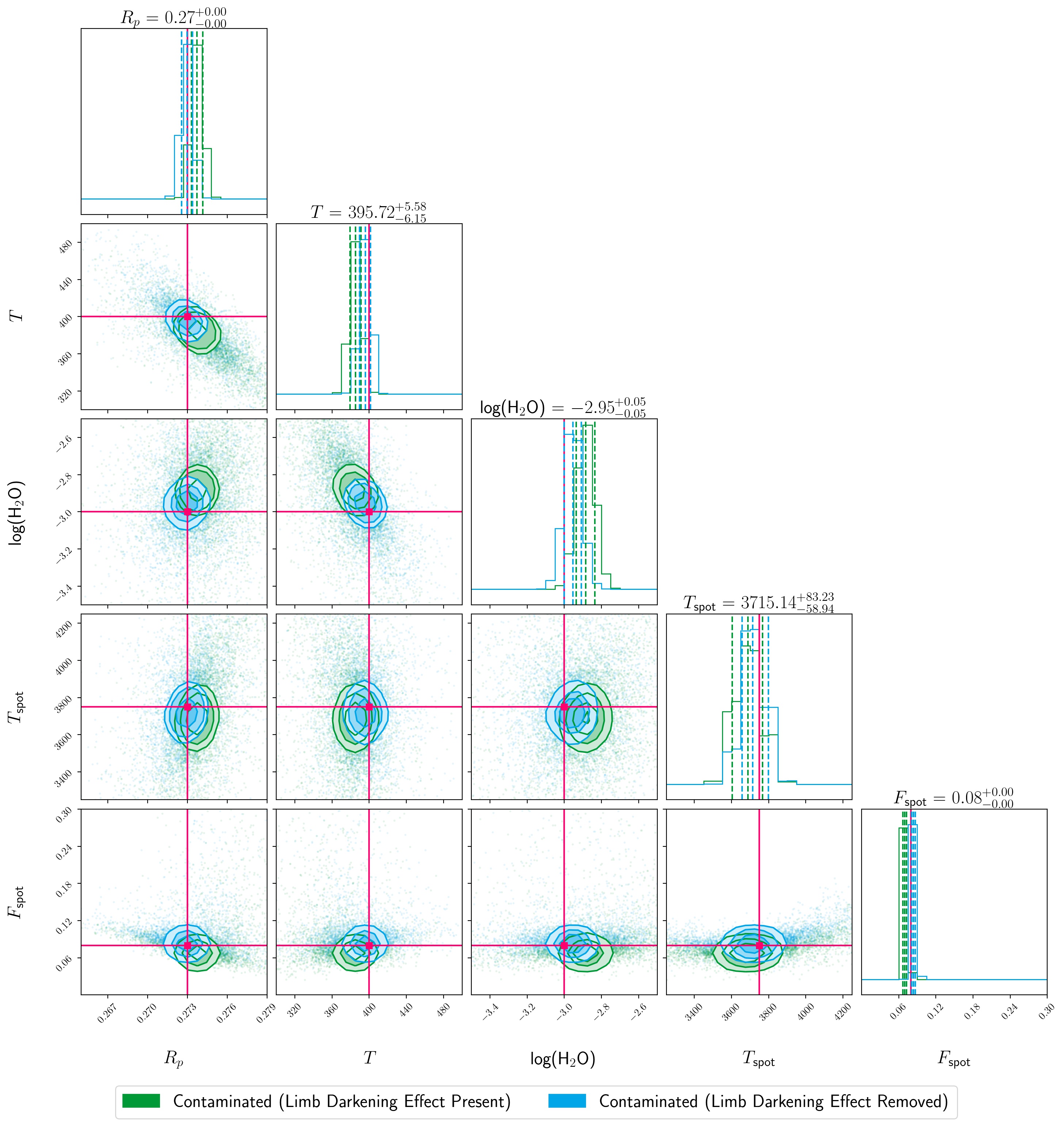}
    \caption{Retrieved posterior distributions for spot Case 9. Figure elements are the same as those in Fig. \ref{fig:case7_posteriors}}
    \label{fig:case9_posteriors}
\end{figure}

\clearpage
\section{Posterior Distribution for the Multiple Spot Cases}
\label{appendixB}

\begin{figure}[htp]
    \centering
    \includegraphics[width=0.9\textwidth, height=0.7\textheight]{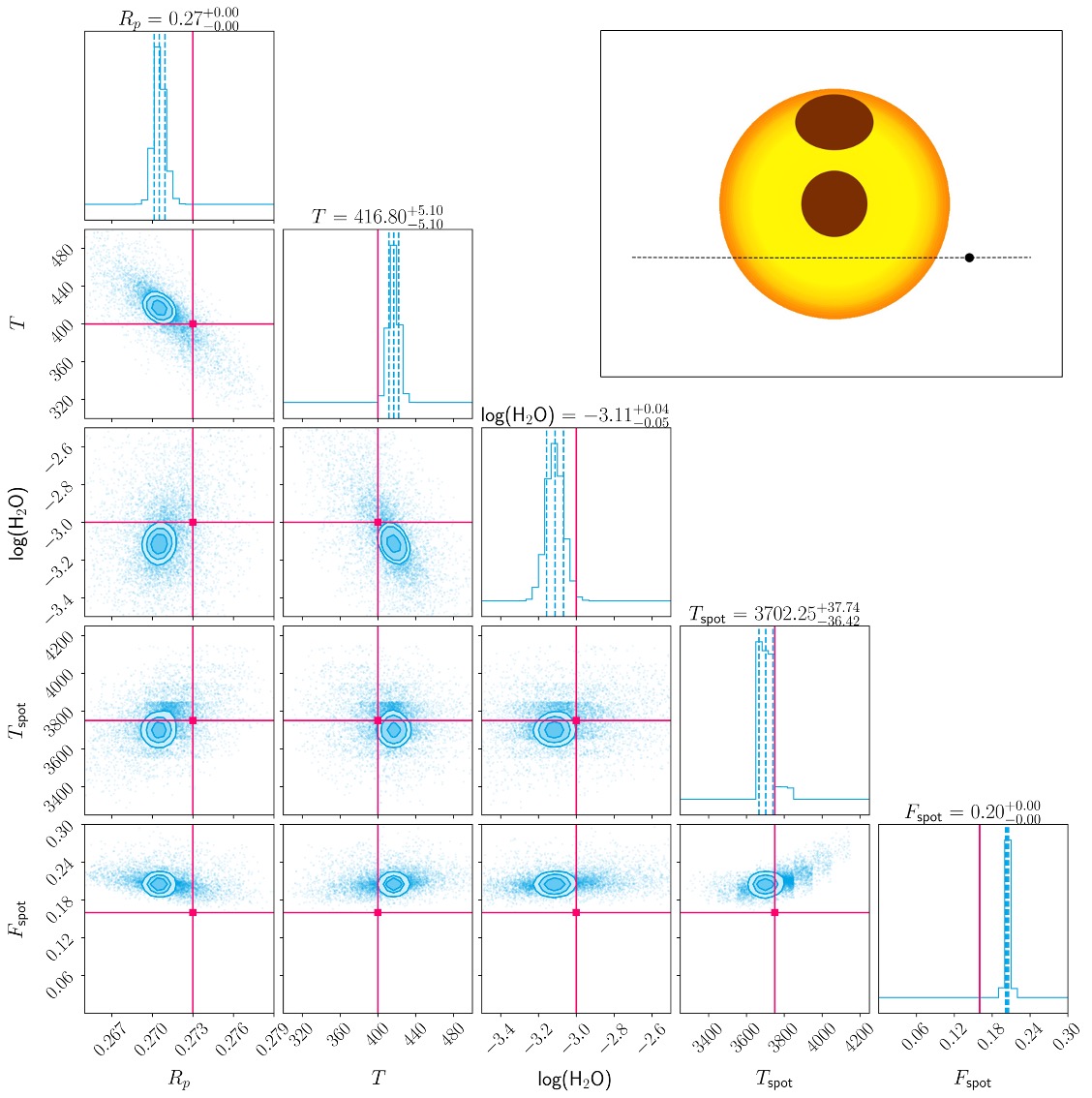}
    \caption{Retrieved posterior distributions for the two spot case (Section \ref{multispot}). The pink lines indicate the ground truth values for each parameter. Inset: a graphical depiction of the stellar disk.}
    \label{fig:twospotposterior}
\end{figure}
\clearpage

\begin{figure}[htp]
    \centering
    \includegraphics[width=0.9\textwidth, height=0.7\textheight]{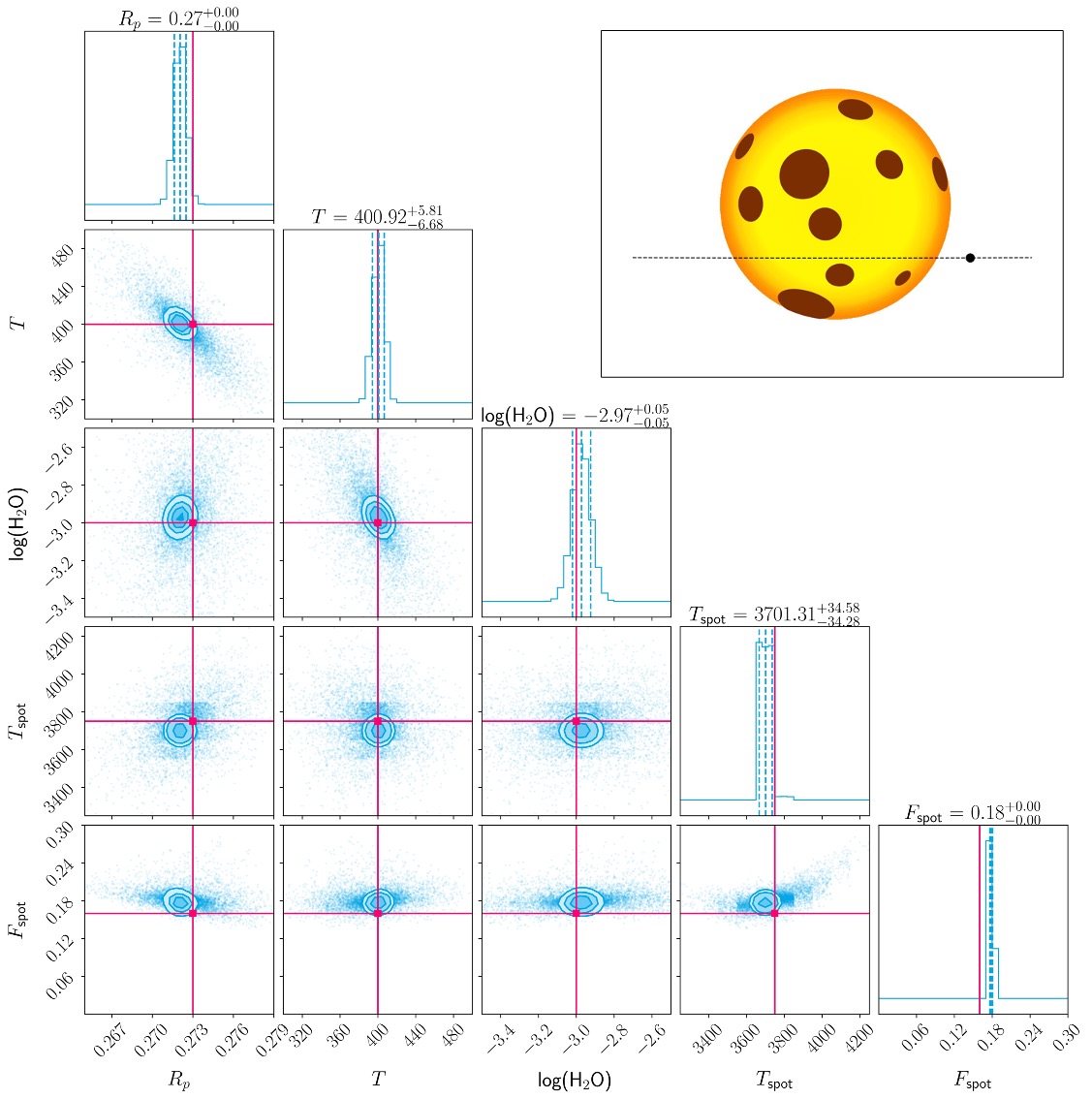}
    \caption{Retrieved posterior distributions for the multiple spot case in which the spots have a random configuration (Section \ref{multispot}). The pink lines indicate the ground truth values for each parameter. Inset: a graphical depiction of the stellar disk.}
    \label{fig:multirandomspotposterior}
\end{figure}
\clearpage

\begin{figure}[htp]
    \centering
    \includegraphics[width=0.9\textwidth, height=0.7\textheight]{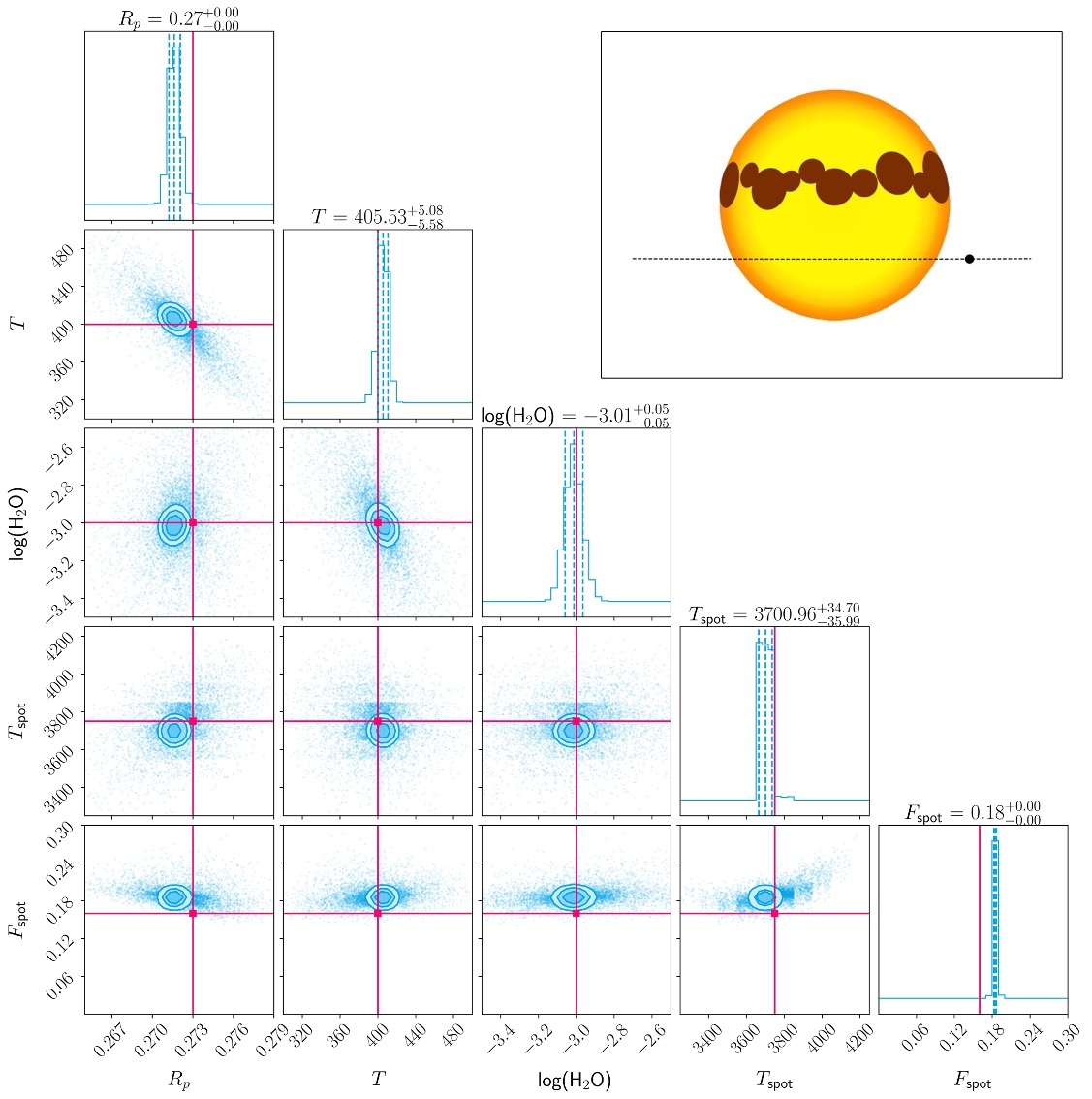}
    \caption{Retrieved posterior distributions for the multiple spot case in which the spots occupy a preferred active latitude centred around $\phi = 15\degree$ (Section \ref{multispot}). The pink lines indicate the ground truth values for each parameter. Inset: a graphical depiction of the stellar disk.}
    \label{fig:activelatposterior}
\end{figure}

\begin{table}[htp]
\centering
\caption{The retrieved planetary and spot parameters for the multiple spot cases considered in Section \ref{multispot} compared to the ground truth values (GT) and Case 7, a single spot of the same temperature and filling factor.}
\label{tab:multispot}
\begin{tabular}{llllll}
\hline
          & $R$ $(R_{Jup})$ & $T$ (K)  & $log(H_{2}O)$ & $T_{Spot}$ (K) & $F_{Spot}$ \\ \hline \hline
GT & 0.273    & 400    & -3       & 3750      & 0.16   \\
Case 7 & 0.2698$_{-0.0005}^{+0.0004}$    & 429.02$_{-5.45}^{+5.98}$ & -3.22$\pm{0.05}$    & 3787.50$_{-59.60}^{+41.23}$   & 0.234$_{-0.006}^{+0.002}$ \\
Two Spot & 0.2705$\pm{0.0004}$   & 416.80$\pm{5.10}$ & -3.11$_{-0.05}^{+0.04}$    & 3702.22$_{-36.41}^{+37.76}$   & 0.203$\pm{0.002}$\\
Multiple Spot - Random & 0.2721$\pm{0.0004}$ & 400.92$_{-6.68}^{+5.81}$ & -2.97$\pm{0.05}$ & 3701.30$_{-34.57}^{+34.30}$ & 0.178$\pm{0.002}$ \\
Multiple Spot - Active Latitude & 0.2716$\pm{0.0004}$ & 405.52$_{-5.58}^{+5.07}$ & -3.01$\pm{0.05}$ & 3700.95$_{-34.57}^{+36.00}$ & 0.185$\pm{0.002}$ \\
\hline

\end{tabular}
\end{table}
\clearpage

\section{Posterior Distributions for the Radial Temperature Variation Spot Cases}
\label{appendixC}

\begin{table}[htp]
\centering
\caption{The input umbra and penumbra parameters for the two radial temperature variation spot cases considered in Section \ref{multitempspot}.}
\label{tab:tempvarinputs}
\begin{tabular}{lllllll}
\hline
& $T_{Umbra}$ (K) & $R_{Umbra}$ ($R_{*}$) & $F_{Umbra}$ & $T_{Pen}$ (K) & $R_{Pen}$ ($R_{*}$) & $F_{Pen}$ \\ \hline \hline
Temperature Variation Case 1 & 3750     & 0.3      & 0.09 & 4250     & 0.4       & 0.07 \\
Temperature Variation Case 2 & 3750     & 0.2      & 0.04 & 4250     & 0.4       & 0.12\\ \hline
\end{tabular}
\end{table}

\begin{table}[htp]
\centering
\caption{The retrieved planetary and spot parameters for the two radial temperature variation spot cases considered in Section \ref{multitempspot} compared to the highest activity, single temperature spot case (Case 7) and the ground truth values (GT).}
\label{tab:tempvarretrieved}
\begin{tabular}{llllll}
\hline
& $R$ $(R_{Jup})$ & $T$ (K)  & $log(H_{2}O)$ & $T_{Spot}$ (K) & $F_{Spot}$ \\ \hline
GT  & 0.273     & 400    & -3       & -         & -     \\
Case 7  & 0.2698$_{-0.0005}^{+0.0004}$    & 429.02$_{-5.45}^{+5.98}$ & -3.22$\pm{0.05}$    & 3787.50$_{-59.60}^{+41.23}$   & 0.234$_{-0.006}^{+0.002}$ \\
Temperature Variation Case 1 & 0.2721$_{-0.0003}^{+0.0004}$    & 446.52$_{-5.66}^{+5.57}$ & -3.39$\pm{0.04}$    & 4201.09$_{-34.46}^{+33.84}$   & 0.254$\pm{0.003}$ \\
Temperature Variation Case 2 & 0.2715$_{-0.0004}^{+0.0003}$    & 440.65$_{-5.84}^{+5.86}$ & -3.33$_{-0.04}^{+0.05}$    & 4299.21$_{-33.68}^{+33.98}$   & 0.278$\pm{0.004}$   \\ \hline
\end{tabular}
\end{table}

\begin{figure}[htp]
    \centering
    \includegraphics[width=0.9\textwidth, height=0.77\textheight]{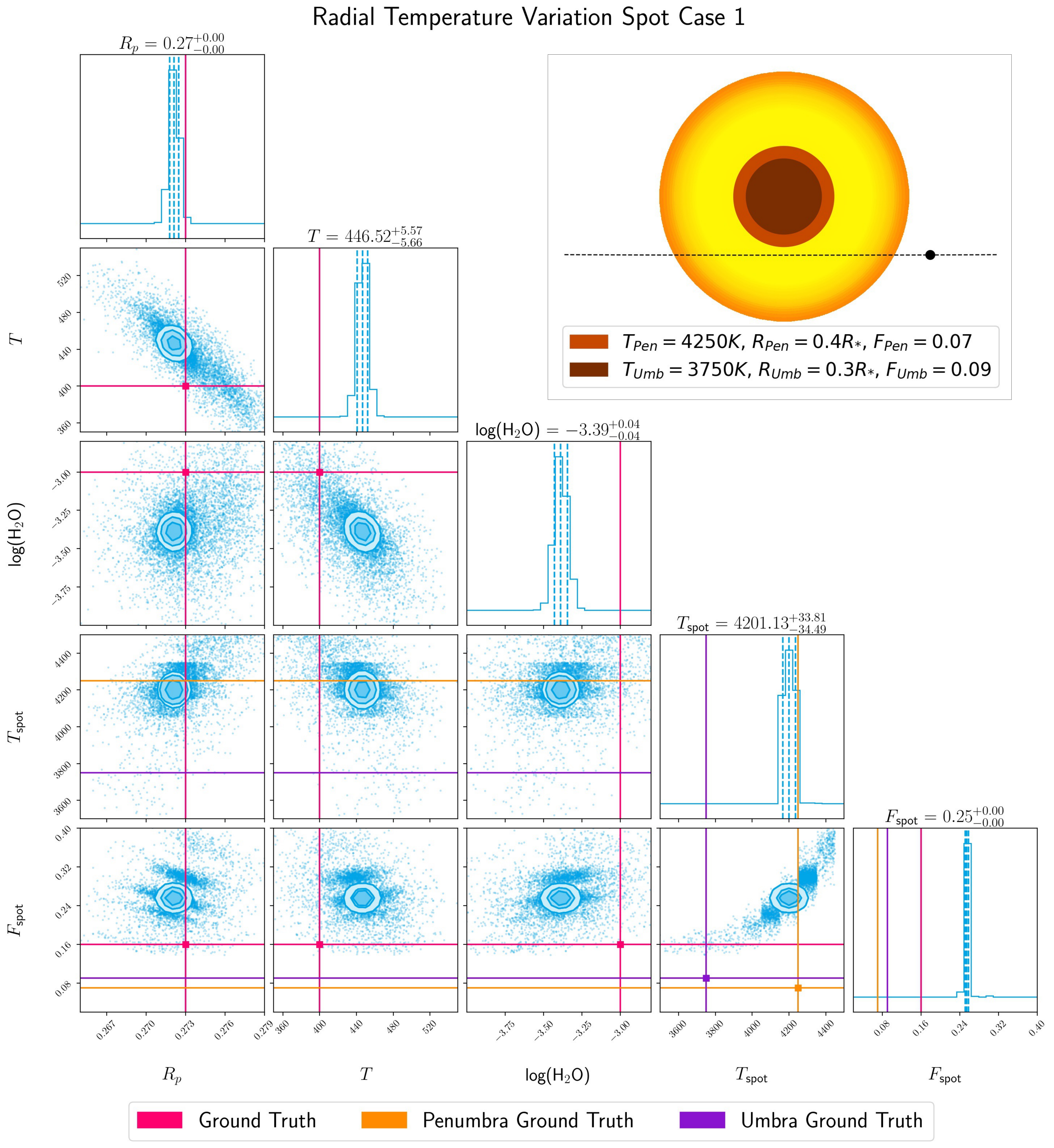}
    \caption{Retrieved posterior distributions for Case 1 when considering a spot with radial temperature variations (Section \ref{multitempspot}). The pink lines indicate the ground truth values for each parameter. The orange and purple lines depict the ground truth values for the spot penumbra and umbra respectively. Inset: a graphical depiction of the stellar disk for this case showing the separation of the spot into an umbra and penumbra and their respective parameters.}
    \label{fig:TVar1_posteriors}
\end{figure}

\begin{figure}[htp]
    \centering
    \includegraphics[width=0.9\textwidth, height=0.77\textheight]{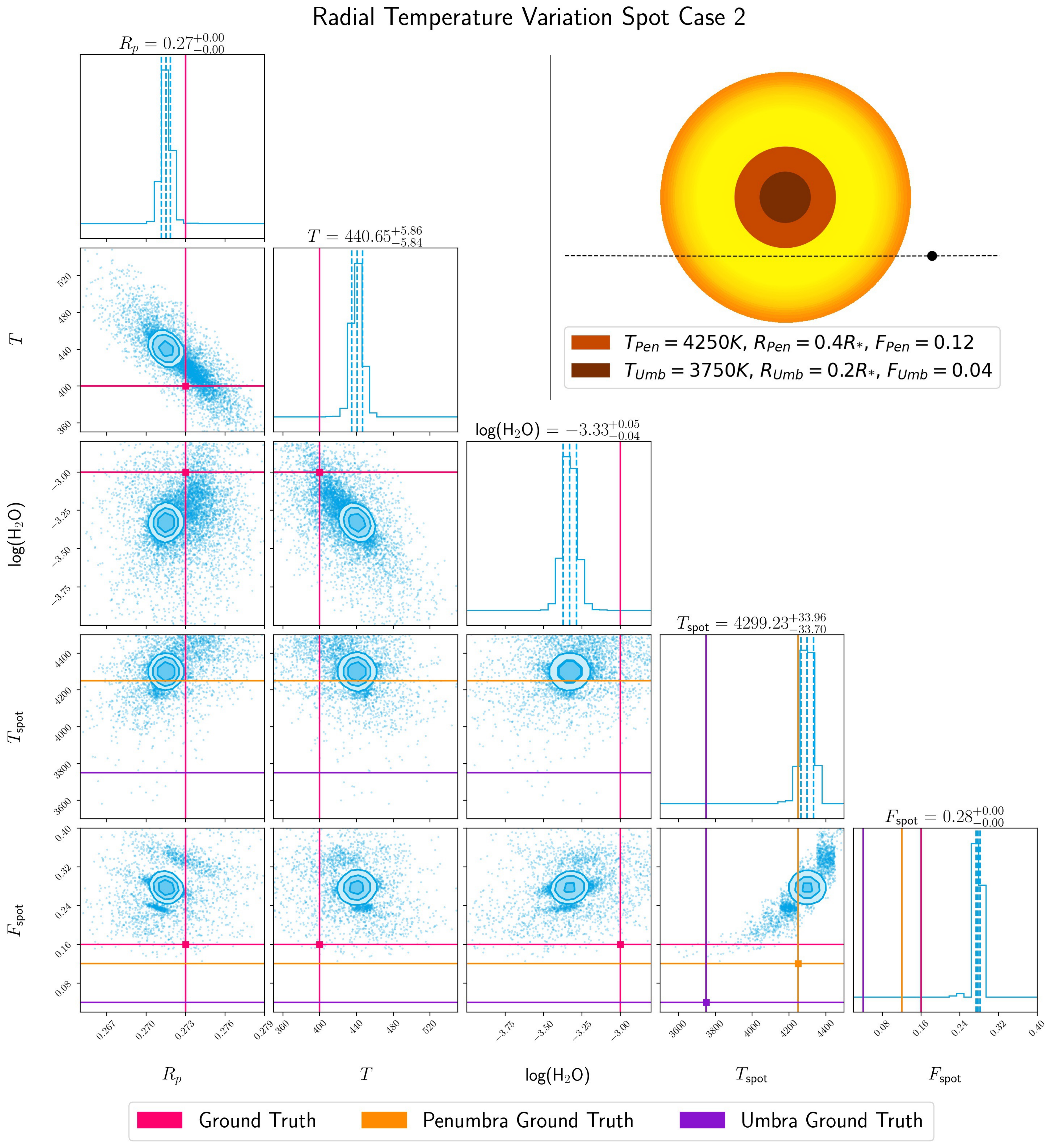}
    \caption{Retrieved posterior distributions for Case 2 when considering a spot with radial temperature variations (Section \ref{multitempspot}). Figure elements are the same as those in Fig. \ref{fig:TVar1_posteriors}.}
    \label{fig:TVar2_posteriors}
\end{figure}

\end{document}